\documentclass[12pt]{article}   
\usepackage{osajnl} 
\usepackage{graphicx}
\usepackage{endfloat}
\usepackage{color}
\usepackage{slashbox}
\usepackage{multirow}
\usepackage{amsmath, amsthm, amssymb}
\numberwithin{equation}{section}

\begin{document}
\title{Calibration and Pre-Compensation \\ of Non-Common Path Aberrations\\ 
  for extreme Adaptive Optics}


\author{J.-F. Sauvage }
\address{Office National d'Études et de Recherches
        Aérospatiales\\ Département d'Optique Théorique et Appliquée\\ BP 72,
        F-92322 Châtillon cedex, France}

\author{T. Fusco}
\address{Office National d'Études et de Recherches
        Aérospatiales\\ Département d'Optique Théorique et Appliquée\\ BP 72,
        F-92322 Châtillon cedex, France}

\author{G. Rousset}
\address{LESIA - Universit\'e Paris 7, \\ Observatoire de Paris, 5 place J.
  Janssen, \\92195 Meudon cedex}

\author{C. Petit}
\address{Office National d'Études et de Recherches
        Aérospatiales\\ Département d'Optique Théorique et Appliquée\\ BP 72,
        F-92322 Châtillon cedex, France}

\email{jean-francois.sauvage@onera.fr}
 
\begin{abstract}Non-Common  Path  Aberrations  (NCPA)  are  one  of  the  main
  limitations for extreme Adaptive Optics (AO) system. NCPA prevent extreme AO
  systems to achieve their  ultimate performance. These static aberrations are
  unseen by the wave front sensor  and therefore not corrected in closed loop.
  We present experimental results validating new procedures of measurement and
  pre-compensation  of the  NCPA on  the AO  bench at  ONERA.  The measurement
  procedure  is  based on  new  refined  algorithms  of phase  diversity.  The
  pre-compensation  procedure makes  use  of a  pseudo-closed  loop scheme  to
  overcome the AO wavefront  sensor model uncertainties. Strehl Ratio obtained
  in  the images  reaches 98.7  \% @  632.8 nm.  This result  allows us  to be
  confident of  achieving the challenging performance  required for extrasolar
  planet direct observation.
\end{abstract}


\maketitle

\section{Introduction}
\label{sect-intro}

Exoplanet direct imaging is one of the leading goal of today's astronomy. Such
a challenge with  a ground-based telescope can only be tackled  by a very high
performance  Adaptive  Optics  (AO)  system,  so-called eXtreme  AO  (XAO),  a
coronagraph      device,      and       a      smart      imaging      process
\cite{Beuzit-p-05,Cavarroc-a-05}.  Most of the  large telescopes  are nowadays
equipped with  AO systems able to  enhance their imaging performance  up to the
diffraction  limit.  One  of  the   limitations  of  the  existing  AO  system
performance  remains  the unseen  Non-Common  Path  Aberrations (NCPA).  These
static  optical aberrations  are  located  after the  beam  splitting, in  the
Wave-Front Sensor (WFS) path and in the imaging path.

The  correction of  NCPA is  one of  the critical  issues to  achieve ultimate
system performance  \cite{Fusco-a-06b} for XAO.  These aberrations have  to be
measured by  a dedicated  WFS tool, judiciously  placed in the  imaging camera
focal plane, and then directly  pre-compensated in the closed loop process. An
efficient way  to obtain such a calibration  is to use a  Phase Diversity (PD)
algorithm    \cite{Gonsalves-82,Paxman-92a,Meynadier-a-99}   for    the   NCPA
measurement.

For  the correction,  the Wave-Front  (WF) references  of the  AO loop  can be
modified to account for these unseen aberrations in the AO compensation and to
directly  obtain  the  best  possible  wave-front quality  at  the  scientific
detector \cite{Blanc-a-03b}.

This  type   of  approach  has   been  successfully  applied   on  NAOS-CONICA
\cite{Rousset-p-02a} or  Keck \cite{vandam-a-04} and has led  to a significant
gain  of  global  system  performance  \cite{Hartung-a-03}.  Even  if  a  real
improvement can  be seen  on pre-compensated images,  a significant  amount of
aberrations are  still not  corrected. In the  frame of the  SPHERE instrument
development \cite{Beuzit-p-05}, we propose  an optimized procedure in order to
significantly   improve   the  efficiency   of   the   NCPA  calibration   and
pre-compensation for high contrast imaging.  The goal is to achieve a residual
Wave-Front  Error  (WFE)  NCPA contribution  of  less  than  10 nm  RMS  after
pre-compensation.

The  principle   of  the  conventional  procedure,  as   used  in  NAOS-CONICA
\cite{Blanc-a-03a,Hartung-a-03},   is  recalled   and  commented   in  Section
\ref{sec-princ}. The  newly optimized algorithm  for the NCPA  measurements is
described  in Section \ref{sec-opt}  and is  based on  a Maximum  A Posteriori
approach  (MAP) \cite{Conan-a-98,Mugnier-a-04} for  the phase  estimation. The
new  approach   for  the  NCPA   pre-compensation  is  presented   in  Section
\ref{sec-opt-precomp}. The application of the PD by the DM itself is discussed
in  Section   \ref{sec-PD-DM}.  In   Section  \ref{exp-res},  we   detail  the
experimental results  obtained with the ONERA  AO bench for  the validation of
the  key  points  of   the  proposed  NCPA  calibration  and  pre-compensation
procedure.

\section{Principle of the NCPA calibration and pre-compensation}
\label{sec-princ}

\subsection{Phase diversity for NCPA calibration}

In  order to  directly  measure the  wave front  errors  at the  level of  the
scientific detector,  a dedicated WFS  has to be  implemented. The idea  is to
avoid  any  additional  optics.  The  WFS  must therefore  be  based  on  the
processing of the focal plane images recorded by the scientific camera itself.
The   phase   diversity    (PD)   approach   \cite{Gonsalves-82,   Paxman-92a,
  Meynadier-a-99, Blanc-a-03b} is a  simple and efficient candidate to perform
such a  measurement. In this section we  are going to briefly  describe the PD
concept and its interest with respect to our particular problematic.

\begin{figure}[htbp]
\begin{center}\leavevmode
  \includegraphics[width=\linewidth]{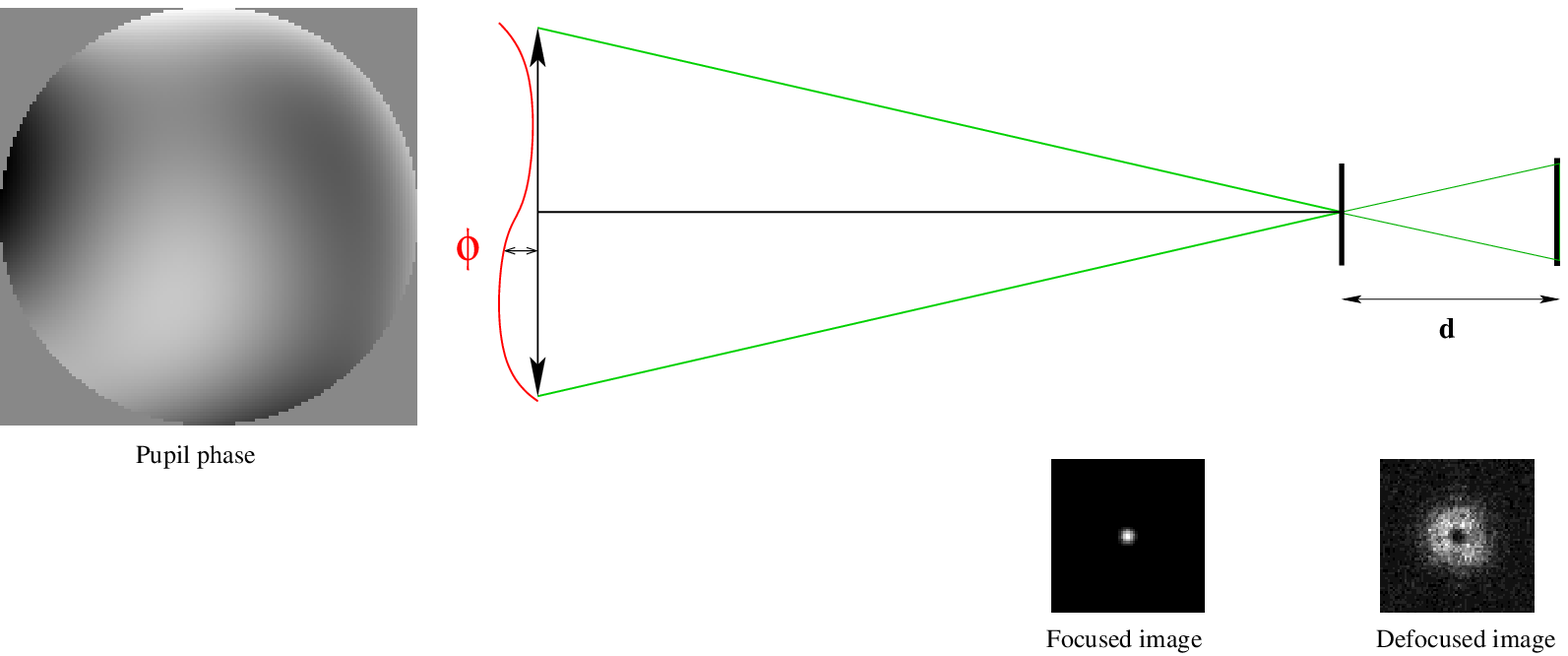}
  \caption{Principle  of phase  diversity :  two images  differing by  a known
    aberration (here defocus).}
  \label{fig-phase_diversity_ppe}
  \end{center}
\end{figure}

The    principle     of    PD   (as     shown    in    Figure
\ref{fig-phase_diversity_ppe}) is to use two focal plane images differing by a
known  aberration  (for instance  defocus), in  order to  estimate the
aberrated phase.

As  shown in  Equation \ref{eqn-i},  the two  images recorded  on  the imaging
camera are  nothing but  the convolution of  the object  and the PSF  (the PSF
being related  to the  pupil phase $\phi$)  plus photon and  detector additive
noises: 

\begin{eqnarray}
  i_f&=&\left|FT^{-1}\left(P \exp(j\phi)\right)\right|^2*o+n  \nonumber \\ 
  i_d&=&\left|FT^{-1}\left(P \exp(j(\phi+\phi_d))\right)\right|^2*o+n
\label{eqn-i}
\end{eqnarray}

where $i_f$ is the conventional  image, $i_d$ the PD image, $j=\sqrt{-1}$, $P$
the pupil function,  $\phi$ the unknown phase, $\phi_d$  the known aberration,
$o$ the observed  object, $n$ the total noise, $*$  stands for the convolution
process, and $FT$ stands for  Fourier Transform. The phase $\phi(\vec{\rho})$
is generally  expanded on  a set  of basis modes  ($\vec{ \rho}$  the position
vector in the pupil plane). Using  the Zernike basis to describe the aberrated
phase we can write:

\begin{equation}
  \phi(\vec{ \rho})=\Sigma_k   a_k \cdot Z_k(\vec{ \rho})
  \label{eqn-expZern}
\end{equation}

where $a_k$ are the Zernike coefficients  of the phase expansion and $Z_k$ the
Zernike  polynomials. The  number of  $a_k$ used  to describe  the  phase will
depend  on   the  performance  required   and  on  the   SNR  characteristics.
Nevertheless,  since optical  aberrations are  considered, only  first Zernike
(typically between 10 to 100) are enough to well describe the NCPA.

As shown in Equation \ref{eqn-i},  there is a nonlinear relation between $i_f$
and $\phi$ (and  thus the $a_k$ coefficients). The estimation  of $\phi$ has to
be     solved    by     the    minimization     of    a     given    criterion
\cite{Meynadier-a-99,Blanc-a-03b}. We  propose hereafter to  define an optimal
criterion adapted to the instrumental conditions (noise and WF aberrations) by
using a MAP approach.

The  most simple known  aberration to  apply is  a defocus~\cite{Gonsalves-82},
(its amplitude is typically of the order of $\lambda$) which can be introduced
in several ways :
\begin{itemize}
\item by translating the camera  detector itself along the optical axis. The
  drawback  of this approach  is the  range of  displacement required  for the
  detector,   especially  when  high   F  ratio   are  considered.   But  this
  theoretically is  the best option  if it can  be implemented. All  the other
  options, presented  below, may  introduce not only  defocus, but  other high
  order aberrations  of very  low amplitudes, especially  spherical aberration
  ($Z_{11}$).  Even if  they are  easily  quantified by  using optical  design
  software, they may, in fine, limit the accuracy of the calibration.
\item by translating a pinhole source in the entrance focal plane of the camera. 
This option has been used to calibrate the NCPA on 
CONICA~\cite{Blanc-a-03a,Hartung-a-03}.
  \item by defocusing the image of a pinhole source in the entrance focal plane 
of the camera by translating the upstream collimator, for instance.
\item and last but not least by using the DM directly. An 
adequate application of a set of voltages on the DM allows us to introduce 
the wanted defocus with an accuracy related to the DM fitting capability.
\end{itemize}
The two main advantages of the DM option are: 
\begin{itemize}
\item no additional optical device is installed in the instrument, e.g. a software 
procedure may be developed to properly offset the voltages of the DM,
\item other types of aberrations, like astigmatism for instance, can also be 
considered leading to a high flexibility in the procedure. Moreover, introducing
other aberrations allows more accurate estimation of the focus itself as
presented in section \ref{sec-PD-DM}.  
\end{itemize}

The DM option was first used in NAOS-CONICA\cite{Blanc-a-03a,Hartung-a-03} and
has also been applied to  Keck\cite{vandam-a-04}. A number of limitations have
been identified in the PD\cite{Blanc-a-03a,Hartung-a-03}: photon and detector
noise, detector defects (flat field  stability), accuracy of the PD (amplitude
and possibly  additional high order aberrations)  and algorithm approximations
\cite{Blanc-a-03a}  (including  the  number  of  Zernike).  In  the  optimized
procedure  we  propose  hereafter,   the  phase  estimation  is  performed  by
minimizing a  Maximum A Posteriori criterion accounting  for non-uniform noise
model   and   phase   \emph{a   priori}   in  a   regularization   term   (see
\cite{Mugnier-a-04}  for  a  detailed  explanation  of  this  approach).  This
optimized PD algorithm is presented in section \ref{sec-opt}.

\subsection{NCPA pre-compensation in the AO loop}
\label{sec-comp_ppe}
\begin{figure}[htbp]
\begin{center}\leavevmode
  \includegraphics[width=\linewidth]{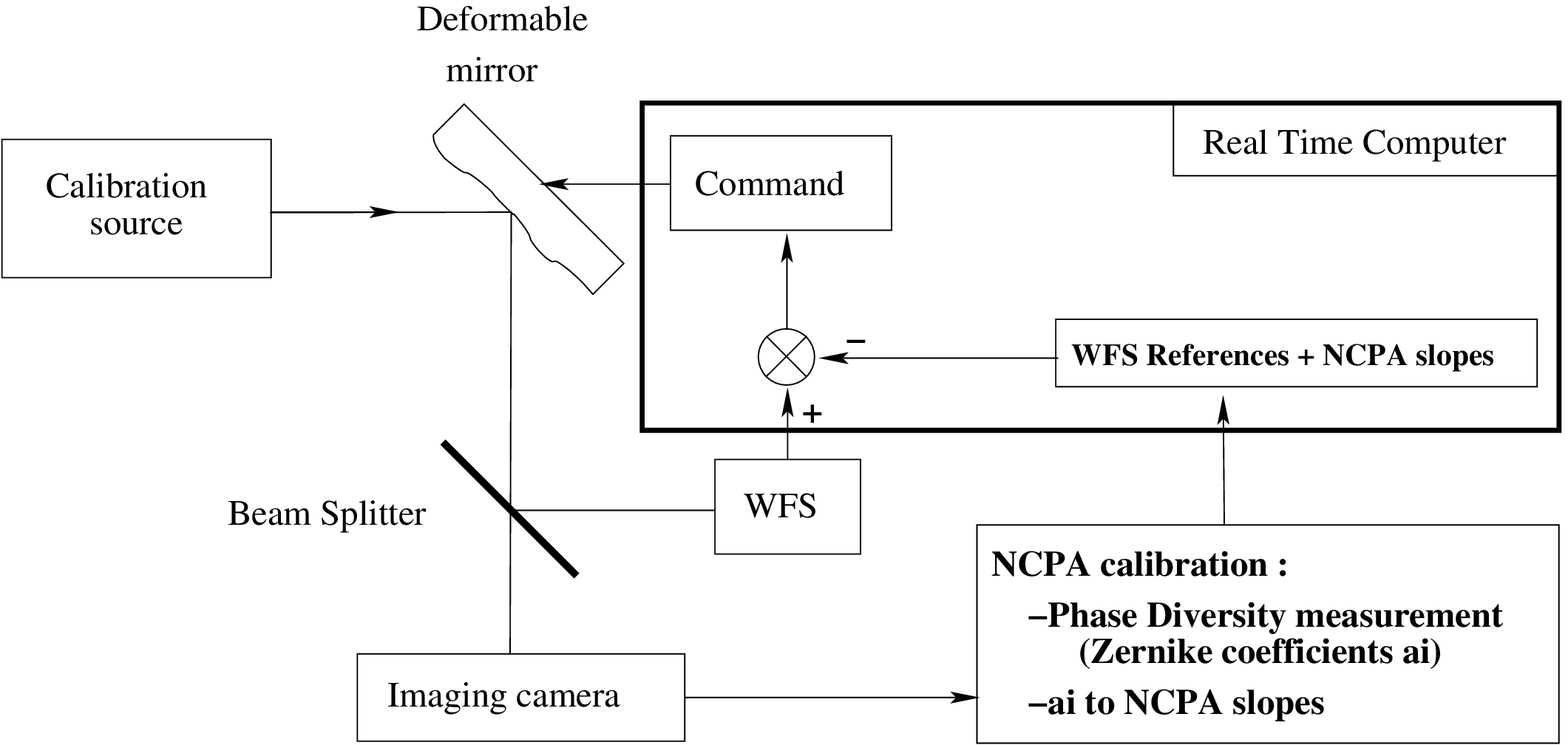}
  \caption{Principe of  NCPA pre-compensation.}
  \label{fig-comp_ppe}
  \end{center}
\end{figure}

The  principle   of  the  NCPA   pre-compensation  is  presented   in  Figure
\ref{fig-comp_ppe}. It consists in modifying the reference of the WFS to
deliver a pre-compensated wave-front to the scientific path. 

A two  steps process  is therefore considered.  Reference slopes  are computed
from PD  data using  a WFS model\cite{Rousset-p-93a}.  For that, a  NCPA slope
vector, as would be measured by the AO WFS (a Shack Hartmann for instance), is
first computed off-line from the  PD-measured set of Zernike coefficients by a
matrix multiplication.

The new  reference vector  is then  added to the  current WFS  reference. Then
closing the AO  loop on the reference  allows us to apply the  opposite of the
NCPA to the  DM. This leads to the compensation for  the scientific camera
aberrations (in addition to the  turbulence) and the enhancement of the image
quality at the level of its detector.

Any error  in the  WFS  model directly affects  the reference modifications,
computed  from the  measured  NCPA, and  thus  limits the  performance of  the
pre-compensation process.  Model errors have  been identified as  an important
limitation of  the approach in NAOS-CONICA\cite{Hartung-a-03}.  As an example,
an error of 10\% on the pixel scale of the AO WFS detector directly translates
into a  10\% uncorrected amplitude  of the NCPA.  One way to  reduce these
model  errors  is    to  perform  accurate  calibrations of  the  WFS
parameters. Nevertheless,  uncertainties on calibrations (pixel  scale of WFS,
pupil  alignment)  will  always   degrade  the  ultimate  performance  of  the
pre-compensation  process. In  order to  overcome this  problem, a  new robust
approach is proposed in Section \ref{sec-opt}.

An important parameter in the  NCPA pre-compensation is the selected number of
Zernike to  be compensated by the  DM. This number  is in fact limited  by the
finite number  of actuators of  the DM, i.e.  the finite number of  degrees of
freedom  of the  AO system.  The DM  can not  compensate for  all  the spatial
frequencies in the aberrant phase. In addition, the actuator geometry does not
really fit properly the spatial behavior of the Zernike polynomials. All these
problems are translated in fitting  and aliasing errors on the compensated WF.
These limitations have  to be taken into account in  the implementation of the
pre-compensation   procedure.    It   will   be   discussed    later   on   in
Section~\ref{exp-res}.

\section{Optimization of the NCPA measurement}
\label{sec-opt}

\subsection{Optimization of the PD algorithm}

As shown in Equation \ref{eqn-i}, there  is no linear relation between $i$ and
$\phi$. Therefore, the estimation of $\phi$ requires the iterative minimization
of a given  criterion. We propose here to define  an optimal criterion adapted
to our experimental conditions (noise and phase to
estimate) by using a MAP approach \cite{Conan-a-98,Mugnier-a-04}.

The MAP criterion is based  on a bayesian scheme (see Equation \ref{eqn-prob})
in which one wants  to maximize the probability of having   object $o$ and 
phase $\phi$  knowing the  images $i_f$ and  $i_d$. 
\begin{equation}
  P(o,\phi | i_f,i_d)=\frac{P(i_f,i_d|o,\phi)P(o)P(\phi)}{P(i_f)P(i_d)}
  \label{eqn-prob}
\end{equation}
The  decomposition  of  this  probability  makes  different  terms  appear  as
discussed in detail in the following paragraphs.

The denominator term $P(i_f)P(i_d)$ stands for the probability of obtaining the
images $i_f$ and $i_d$. As the images are already measured, this term is
equal to 1. 

The  term  $P(i_f,i_d|o,\phi)$,  called  ``likelihood term'',  represents  the
probability to obtain the measured  data considering real object and phase. It
is no  more than the  noise statistic  in the image.  The two main  sources of
noise are the detector noise and the photon noise:

\begin{itemize}
\item for high flux pixels in the image, the dominant noise is the photon one.
  Hence, it follows  a Poisson statistical law which can  be approximated by a
  non-uniform Gaussian law with a variance $\sigma_{i}^2(\vec r) \simeq i(\vec
  r)$, as soon as $i(\vec r)$ is  greater than a few photons per pixel. ($\vec
  r$ a position vector in the focal plane)
\item  for low  flux pixels,  the dominant  noise is  the detector  noise,
  described by a spatially uniform distribution (same variance $\sigma_e^2$ 
  for each pixel) and a gaussian statistical law.
\end{itemize}
Therefore,  the global  noise statistics  can be approximated by  a  non-uniform 
gaussian law of variance \cite{Mugnier-a-04} 
$\sigma_{i}^2(\vec r)=\sigma_e^2+i(\vec r)$. 


The second term,  $P(o)$, represents the \emph{a priori}  knowledge we have on
the  object. In  our case,  the object  is marginally  resolved (less  than two
pixels, for a  diffraction FWHM of 4 pixels). Nevertheless  to account for its
small extension  as well as to account  for pixel response, we  have chosen to
consider it as an unknown in the PD process.

In the  following, the only  prior imposed on  the object will be  a positivity
constraint  (using   a  reparametrisation   :  $o=a^2$)  leading   to  
$P(o)=P(a^2)=1$.  The probability  $P(o)$  does not  impact  on the  criterion
minimization. 

\begin{figure}[htbp]
\begin{center}\leavevmode
  \includegraphics[width=\linewidth]{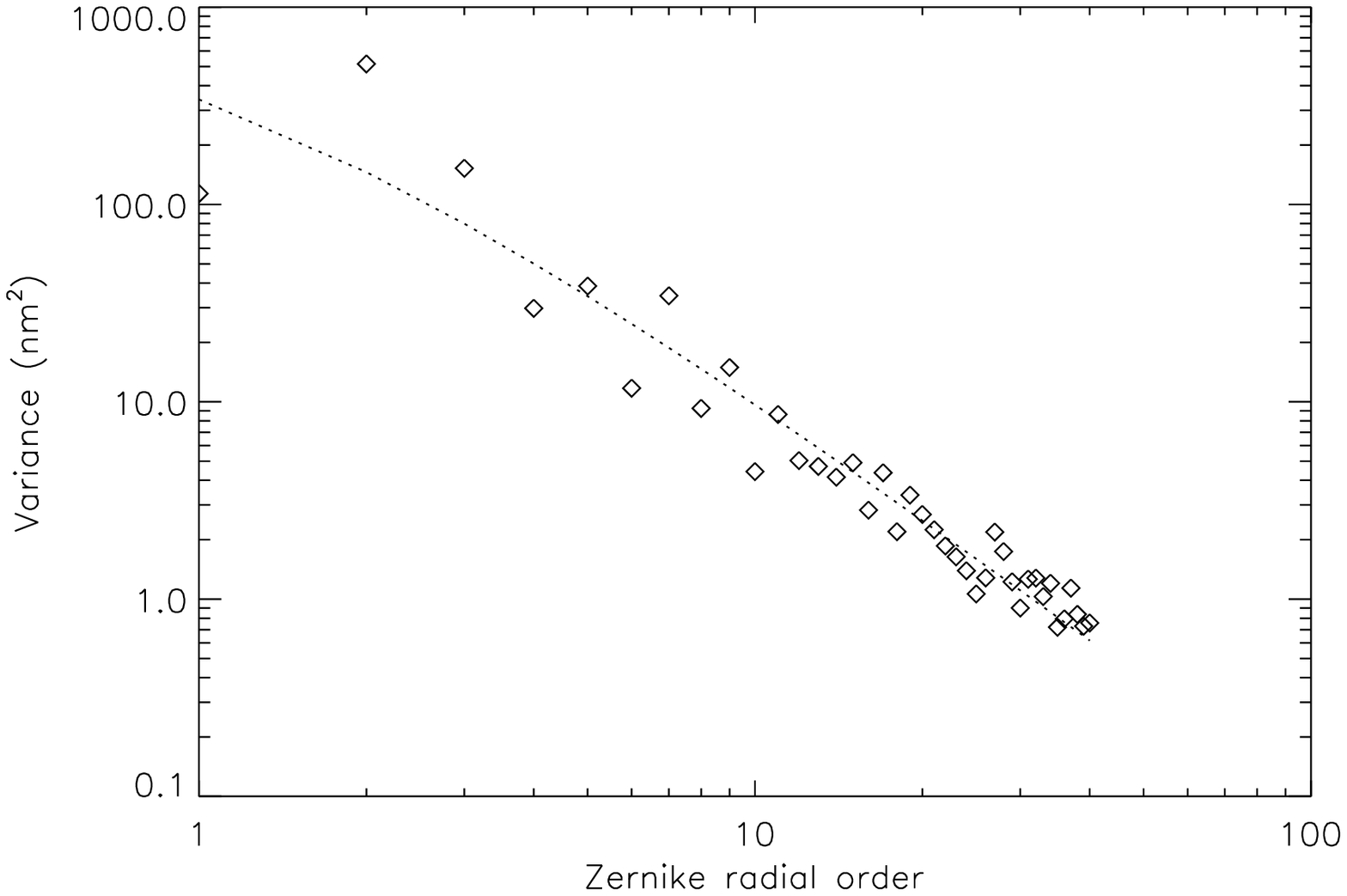}
  \caption{Typical aberration  spectrum measured on  existing optics.  The diamonds  show  the measured
    Zernike coefficient, integrated on radial order and the dotted line shows
    the $(n+1)^{-2}$ approximation.}
  \label{fig-zern}
  \end{center}
\end{figure}

The  third  term,  $P(\phi)$,  is   the  regularization  term  for  the  phase
estimation. It  accounts for the knowledge we  have on the NCPA.  Note that in
our case we  have chosen to parameterize the phase  $\phi$ by the coefficients
$a_k$ of  its expansion on  the Zernike basis.  Then, the term  $P(\phi)$ will
easily  solve the  problem  of  the choice  of  the number  of  Zernike to  be
accounted  for in the  estimation of  the phase.  Let us  mention that  in the
conventional  PD  approach  \cite{Paxman-92a,Meynadier-a-99,Blanc-a-03a},  the
relatively arbitrary choice of a given number of Zernike (the $N$ first) to be
estimated, is in fact an  implicit regularization of the estimation problem by
the truncation of the phase expansion  in order to avoid the noise propagation
on the Zernike  high orders. This corresponds to a  reduction in the dimension
of the  solution space. Here  we will select  in the algorithm  a sufficiently
large number of  Zernike so as to not significantly  reduce the solution space
and  to regularize  the estimation  by the  term $P(\phi)$  in  the criterion.
Indeed the NCPA are composed of  static aberrations due to the optical design,
polishing defects, and misalignments.  In a first approximation, their spatial
power  spectral density  follows a  $(n+1)^{-2}$ law  where $n$  is the
Zernike radial order (see Figure \ref{fig-zern}). A general form for $P(\phi)$
is given by:
\begin{equation}
  P(\phi)=\exp(-\phi^tR_{\phi}^{-1}\phi)
    \label{eqn-p-phi}
\end{equation}
where $R_{\phi}$  is the phase covariance  matrix and has on  its diagonal the
variance of the  Zernike coefficients, of similar behavior  than the one given
in Figure \ref{fig-zern}, the other coefficients (covariances) being put equal
to zero.

The phase  estimation is done  by minimizing $J(o,\phi)$
equal to  $-\ln P(i_f,i_d,o,\phi)$:
\begin{equation}
  J(o,\phi)=\left|\left|\frac{i_f-h_f*o}{\sigma_f(\vec{r})}\right|\right|^2 
     +\left|\left|\frac{i_d-h_d*o}{\sigma_d(\vec{r})}\right|\right|^2
     +\phi^tR_{\phi}^{-1}\phi
  \label{eqn-likelyhood}
\end{equation}
This criterion makes appear the non-uniform noise statistics in the two images
with  standard  deviations  $\sigma_f(\vec{r})$ and  $\sigma_d(\vec{r})$.  The
covariance matrix $R_{\phi}$ represents the  prior knowledge of the phase. The
minimization algorithm  is based on  an iterative conjugate  gradient approach
allowing a fast  convergence \cite{Blanc-a-03b,Mugnier-a-04}. For the starting
guess, all the Zernike coefficients are put to zero. Note that in the previous
algorithm            used            for            the            NAOS-CONICA
calibration~\cite{Blanc-a-03a,Hartung-a-03}, the PD  algorithm did not include
the non-uniform noise statistics and the phase regularization term.


\subsection{Simulation results}
\label{subsec-sim}
We  only present  here  improvements  brought by our new algorithm,   the
non-uniform noise model and  phase regularization, both  in simulation in this
section and experimentally in Section \ref{exp-res}.

Simulation is divided  into two main parts : the  generation of noisy aberrant
images  and the  phase estimation  by different  PD algorithms.  The simulated
images  are  $128\  \times\  128$  pixels, and  generated  according  to  some
realistic  parameters of  the ONERA  AO  bench (see  section \ref{exp-res})  :
oversampling factor of 2.05 (this means 4.1 pixels in the Airy spot FWHM), the
aberrant  phase  is modeled  using  the  200  first Zernike  polynomials.  The
aberrations  have  a  total RMS  error  of 45  nm  and  a spectrum  shape  of
$(n+1)^{-2}$. Finally, the images are noised with a uniform 1.6 electron noise
per pixel and with photon noise. All $a_k$ values are given in nm. The maximum
flux in  the images  is 100  photons in the  case of  the test  of regularized
algorithm.

We define the SNR in the images as the SNR of the focal plane image expressed
by the ratio of the maximum of the image $i_{max}$ in photon-electrons by the
standard deviation of the sole detector noise in electrons:

\begin{equation}
  SNR=\frac{i_{max}}{\sigma_e}
  \label{eqn-snr}
\end{equation}

\subsubsection{Gain brought by non-uniform noise model}

Let us  first study  the gain  brought by accounting  for a  non-uniform noise
model  in  the  PD algorithm.  For  a  given  aberrant phase,  the  efficiency
$\Sigma_{NU}$ quantifies  the gain in estimation accuracy  for the non-uniform
algorithm with respect to the uniform algorithm:

\begin{equation}
\Sigma_{NU}=100 \times \frac{\epsilon_U-\epsilon_{NU}}{\epsilon_U}
\label{eqn-nunm}
\end{equation}

where $\epsilon_U$ and $\epsilon_{NU}$  are the reconstruction errors obtained
respectively with uniform and non-uniform noise models (in nm$^2$) defined by
Equation \ref{eqn-eps}. \\

\begin{equation}
  \epsilon_X=\sum_{i=1}^{N_{max}}(a_{k_{measured,\ X}}-a_{k_{true}})^2 
  \label{eqn-eps}
\end{equation}

where $a_{k_{measured,\  X}}$ are the estimated Zernike  coefficients with the
uniform ($X=U$) noise model or with the non-uniform noise model ($X=NU$), and
$a_{k_{true}}$ are the true coefficients.

\begin{figure}[htbp]
  \begin{center}
     \includegraphics[width=\linewidth]{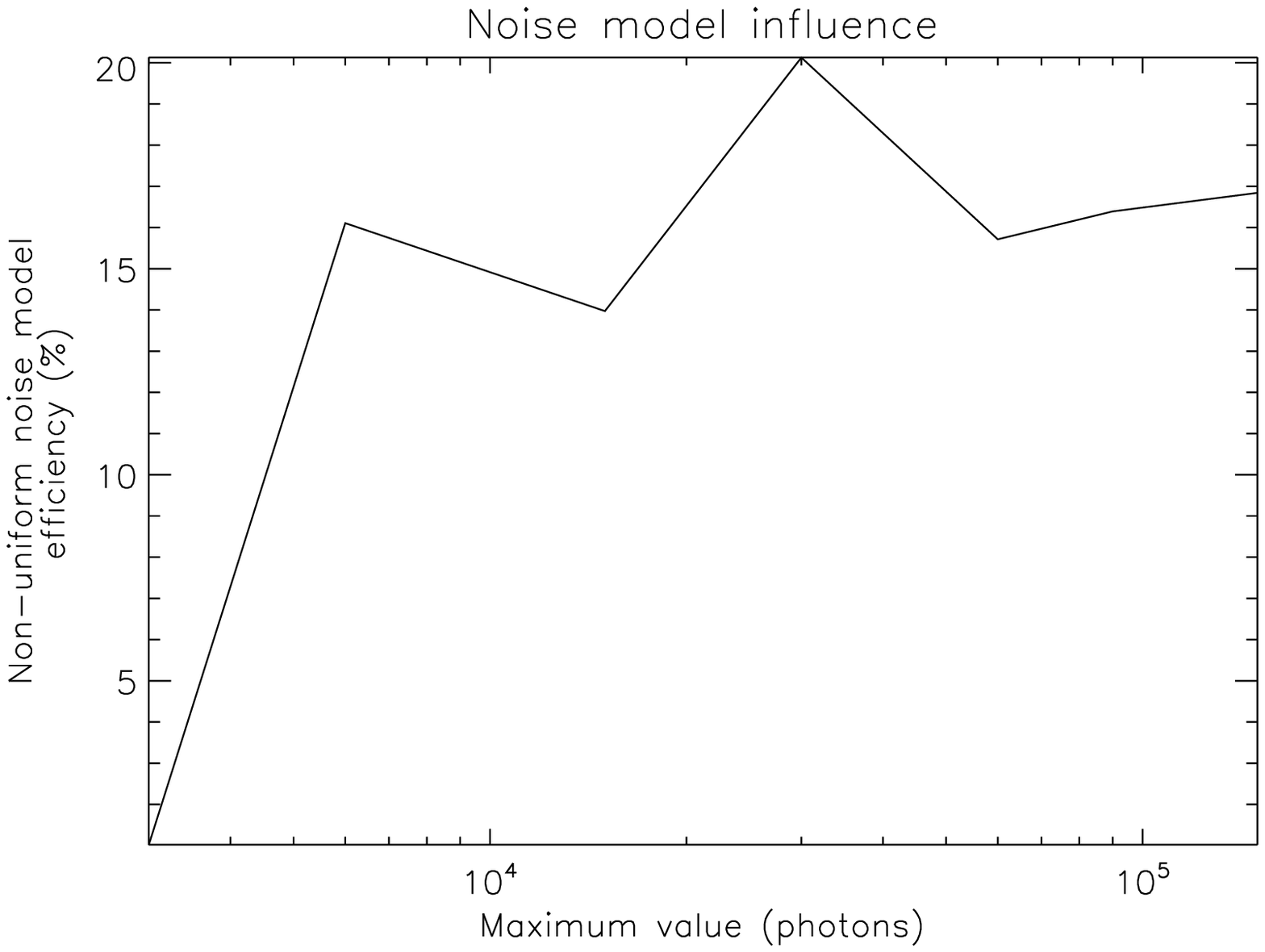}
     \caption{Gain  of  the  non-uniform  model  with  respect  to  conventional
       algorithm  versus the  maximum intensity value  in the  image.  Randomly simulated
       images  of 45nm RMS  aberrant phase,  $(n+1)^{-2}$ shaped  spectrum. 75
       Zernike modes from $Z_4$ to $Z_{78}$ are estimated by PD.}
     \label{fig-non_uni}
  \end{center}
\end{figure}

Figure  \ref{fig-non_uni}  shows the  influence  of  the  noise model  on  the
estimation  accuracy. $\Sigma_{NU}$  is plotted  with respect  to  the maximum
intensity value in the image. Each  point on the curve corresponds to only one
occurrence of noise and aberrations.  Hence the relatively instability found in
computing  $\Sigma_{NU}$.  At low  photon  flux,  both  estimation errors  are
identical. The  image SNR is limited  by detector noise (uniform  noise in the
full image),  and therefore  taking into account  an additive photon  noise is
useless. For high flux, photon noise is predominant and the algorithm with the
non-uniform noise model allows us to increase the phase estimation accuracy of
15 to 20\%  with respect to the  uniform noise model. 

\subsubsection{Gain brought by  phase regularization}



\begin{figure}[htbp]
  \begin{center}\leavevmode
    \includegraphics[width=\linewidth]{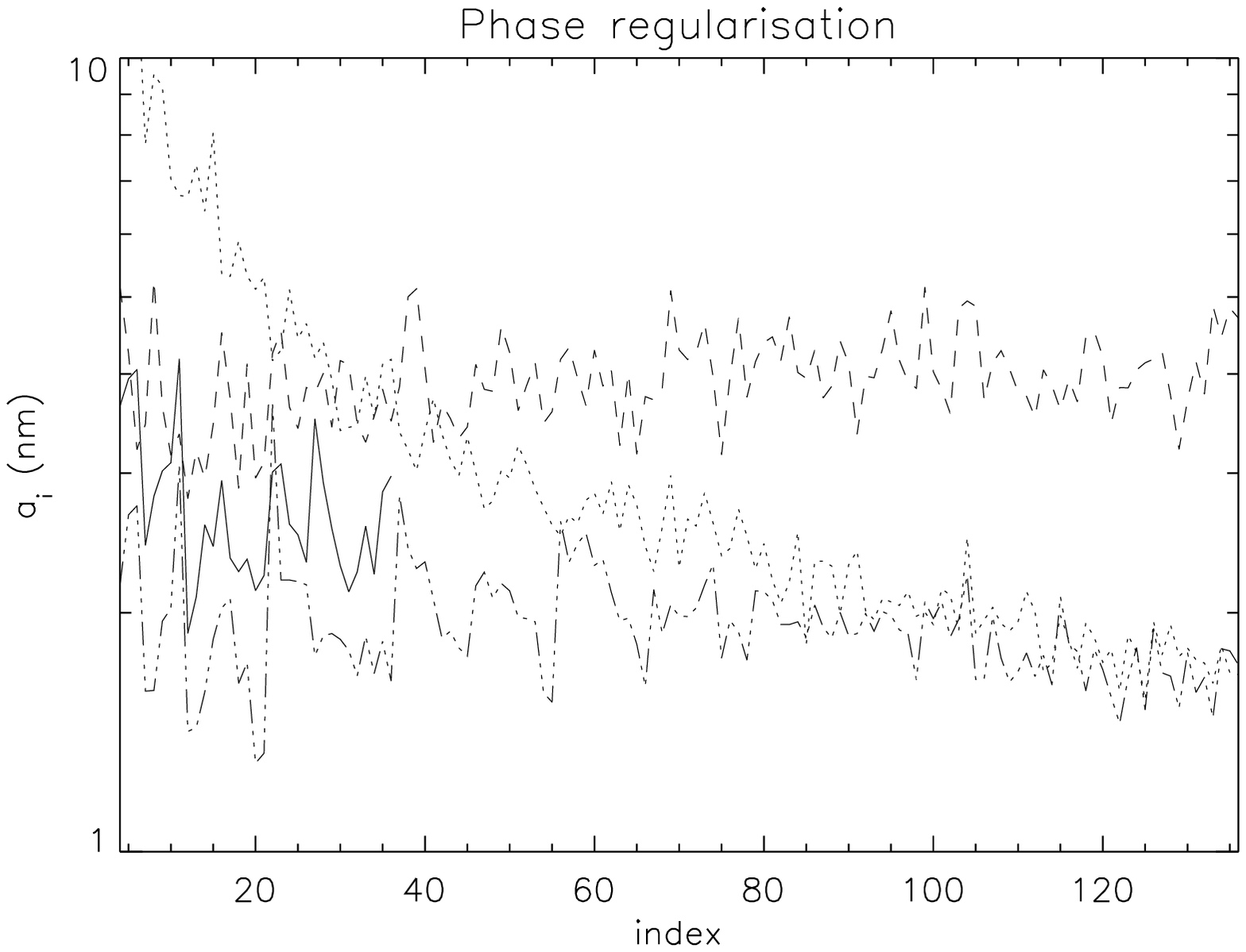}
    \caption{Noise  propagation  in  the  Zernike coefficient  estimation  for
      different  cases  of regularization.  The  dotted  line  shows the  average
      spectrum of  the simulated  aberrations ($45nm$ total  WFE, $(n+1)^{-2}$
       spectrum). The dashed line shows the phase estimation error for a
      133 Zernike estimation, without  any regularization term. The solid line
      shows  the same  estimation,  with only  33  Zernike (regularization  by
      truncation). The dashed-dotted line shows the 133 Zernike estimated with
      the regularization term. SNR in the images is $10^3$.}
    \label{fig-regul_simu}
 \end{center}
\end{figure}

The gain brought  by the phase regularization term  in \ref{eqn-likelyhood} is
quantified in  this section. Figure~\ref{fig-regul_simu}  presents the results
of the simulation using different algorithms. Table \ref{tab-regul_simu} shows
the total  estimation error for each  algorithm, and also  the contribution of
low orders (from $a_4$ to $a_{36}$)  and the contribution of high orders (from
$a_4$ to $a_{137}$) in the total error.

In  dotted line  the estimation  error  in nm$^2$  for the  133 first  Zernike
coefficients (from $a_4$  to $a_{137}$) with a simple  least square estimation
is given.  As comparison, the 133  first coefficients part of  the 200 Zernike
polynomials  simulated from  the  input  spectrum (45nm  RMS)  to compute  the
images,  are plotted in  dashed line.  For this  estimation, the  error (noise
propagation) is constant whatever the  coefficient, as predicted by the theory
(see  \cite{Meynadier-a-99}).  For  coefficients  higher  than  $a_{36}$,  the
estimation error becomes  greater than the signal to  be estimated. The signal
to noise  ratio on these coefficients  is lower than one,  their estimation is
not  possible. The  total error  is 46nm  for the  133 coefficients,  which is
similar to the introduced WFE.

In contrary, an estimation of only the 33 first coefficients $a_4$ to $a_{36}$
(dashed-dotted  line) shows  that the  reconstruction error  is  still roughly
constant whatever the  mode, but fainter than in  the previous estimation. The
total  error  is 29nm  for  the 133  coefficients,  considering  that all  the
estimated coefficients  from $a_{37}$  to $a_{137}$ are  equal to  zero. Here,
reducing  the number  of parameters  in  the estimation  (especially the  high
frequencies of the phase) leads to  a better estimation on the low order modes
and a dramatic decrease of  their estimation error. Nevertheless, as explained
before, this estimation is nothing but a first rough regularization and is not
optimal.

Because the previous regularization is arbitrary, we can refine the estimation
by  using  the prior  knowledge  on  the phase  to  be  estimated,  that is  a
$(n+1)^{-2}$ spectrum  as a regularization  term. In that case,  the estimated
133  first coefficients  are  given by  the  dashed-dotted line.  For all  the
coefficients, the  error is lower  than the input coefficients  (dotted line).
More  important, the  total  error (22nm)  is  smaller than  the  one for  the
estimation  on  33  coefficients.  The  error computed  with  only  the  first
coefficients  $a_4$ to  $a_{36}$ is  also fainter  with regularization,  12 nm
compared to  16nm. For high order  modes, the error  tends to be equal  to the
phase itself, no noise propagation. MAP  allows to optimally deal with low SNR
and avoids any noise amplification. With the regularization term.

\begin{table}
\caption{Estimation error with different phase regularizations. Introduced WFE
is given in comparison. All values are given in nanometers. }
\begin{center}
  \begin{tabular}{|c|c|c|c|c|} \hline 
    & Introduced  & Least square & Truncated least square & Phase
      regularization \\ 
    &  WFE & 133 coefficients& 33 coefficients & 133 coefficients \\ \hline
    $a_4$ to $a_{36}$ & 38 & 21 & 16 & 12 \\ \hline
    $a_{37}$ to $a_{136}$ & 24 & 41 & 24 & 19 \\ \hline
    {\bf Total error} & \bf{45} & \bf{46} & \bf{29} & \bf{22} \\ \hline
\end{tabular}
\label{tab-regul_simu}
\end{center}
\end{table}

\section{Optimization of the NCPA pre-compensation}
\label{sec-opt-precomp}

\subsection{Pseudo-closed loop process}

After PD  measurement, the pre-compensation of  NCPA has to be  performed by a
modification  of WFS  reference. The  compensation accuracy  therefore greatly
depends on  AO loop model.  In order to  overcome this problem and  reach much
better performance,  a new approach  is proposed, "Pseudo-Closed  Loop" (PCL).
The idea  is to use a  feedback loop for the  NCPA pre-compensation, including
the  PD estimation (see  the schematic  of Figure  \ref{fig-comp_ppe}). Indeed
after  a first  pre-compensation of  the  NCPA, it  is mandatory  to have  the
capability to  acquire a  new set  of two pre-compensated  images in  order to
quantify the  residual NCPA due  to model uncertainties.  This can be  done by
closing  the  AO  loop  on  the  artificial  source  used  for  the  PD  image
acquisition, accounting for  the new WF reference. This  ensures the stability
of the pre-compensation by the  DM during the image acquisition. By estimating
a new set of Zernike coefficients,  we have access to the residual phase after
correction. We can take advantage  of its measurement to offset the previously
modified WF  reference. The  process can then  be performed  till convergence,
resulting in  quasi null  measurements of the  Zernike coefficients  (at least
free from  any model error). Indeed  any error in  the AO WFS model  will only
result in slowing the convergence of the process. In addition, after the first
pre-compensation,  the recorded  images may  exhibit a  much better  signal to
noise ratio due to the higher  concentration of photons in the central core of
the image, leading to a better estimation of the Zernike coefficients.

The practical implementation of this PCL approach is summarized
here below.  Firstly, we  perform a careful  AO WFS calibration:  its detector
pixel scale, the pupil image  position, and the reference slope vector obtained
using a  dedicated calibration source at  the AO WFS entrance  focal plane. We
are therefore able to adjust the initial AO WFS model. Secondly, an artificial
quasi-punctual source is placed at the entrance of the AO bench and is used to
calibrate  the NCPA.  With  this calibration  source,  the DM-WFS  interaction
matrix is   calibrated and, from the measurements,  a new command matrix
is  computed. Then,  the multi-loop  measurement compensation process  is:
\begin{itemize}
\item[1)] measurement of  NCPA with PD. This step can be summarized as follows :
  \begin{itemize}
  \item[i)] closing  the AO loop using the calibrated AO WFS references and 
  recording of a focused image on the science camera,
\item[ii)] applying the  defocus (using slope modification) and  recording  a
  defocused image  with the AO loop, closed once again,
  \item[iii)] computation of NCPA from this pair of images with the PD algorithm
  \end{itemize}
\item[2)] computation of the incremental slope vector using  the currently measured NCPA 
\item[3)] modification of the AO WFS references (to account for the latest measurements) 
and saving  the new AO WFS references,
\item[4)] measurement of  residual NCPA with PD and closed  AO loop using the
  new references for the pre-compensation, similar to step 1,
\item[5)] repeat steps 2 through 4 until convergence 
\end{itemize}
A  refinement could  be  to re-calibrate  at each  step  (3 to  4) the  DM-WFS
interaction matrix taking into account  the influence of the reference offsets
in the AO  WFS response and re-compute the command matrix  to achieve the best
possible  efficiency with the  AO system.  \\

An  alternative  approach,  recently  proposed,  is  to  directly  perform  an
interaction  matrix linking  the Zernike modes  to be  compensated and  the PD
estimation (\cite{Kolb-p-04}).

\subsection{Number of compensated modes}
\label{sec-number_compensated_modes}

The number of  Zernike modes that can be compensated for  is determined by the
number of actuators of the DM.  The larger the number of actuators, the better
the fit  of Zernike polynomial by the  DM. We performed the  simulation of the
capability  of our  DM  (69 valid  actuators)  to compensate  for the  Zernike
polynomials,  using  the  DM  influence   functions  as  measured  by  a  Zygo
interferometer.  The  results,  not  presented here,  shows  that  considering
Zernike  polynomials of  radial degree  larger  than 6  leads to  significant
fitting errors  (larger than 45\%  of input standard deviation),  reducing the
overall performance of the NCPA  pre-compensation. In most of the experimental
results presented in this paper in  Section \ref{exp-res}, we have used the 25
first Zernike polynomials  (from defocus $Z_{4}$ up to  $Z_{28}$) for the NCPA
compensation.  It allows  us  to  minimize the  coupling  effects between  the
compensated Zernike  due to  the limited  number of actuators  on the  DM. The
compensation of the 25 first  Zernike polynomials brings already a significant
reduction  of the NCPA  amplitudes. Because  of the  expected decrease  of the
amplitude of the  NCPA with the order of  Zernike (see Figure \ref{fig-zern}),
this choice is not an important limitation in the final performance.

\section{DM application of a phase diversity}
\label{sec-PD-DM}
To finalize the discussion on  the procedure to measure and pre-compensate for
the NCPA, let us now consider the  application of the PD by the DM. As already
stated,  we consider  that this  approach  is probably  the best  for a  fully
integrated  AO  system  in an  instrument  if  there  is no  science  detector
translation capability. For  instance, defocus can be introduced  by moving an
optical element on the  sole optical train of the AO WFS  and closing the loop
with    this    aberration,    as    first   implemented    in    NAOS    (see
\cite{Blanc-a-03a,Hartung-a-03}). But implementing a moving optical element is
an issue with an instrument  requiring high stability. Therefore, we propose to
apply the  PD by the modification  of the AO  WFS references, same as  for the
NCPA pre-compensation. This is a pure software procedure which uses the AO WFS
model. Closing the AO loop with the modified references will apply the defocus
to  the  science  camera but  also  ensures  the  stability of  this  defocus.
Application of  the defocus directly  on the DM  voltages and not  closing the
loop will suffer from DM creeping.  In fact, any other low order aberration can
be considered by this method as PD, allowing a maximum flexibility.

Due to  the uncertainty of the  AO WFS model,  the introduced PD will  not be
perfectly known which results in measurement  errors. The main effect is
the uncertainty on the amplitude of the PD.

Considering defocus as  the known aberration in the  simulation, we observed a
linear dependence  of the  NCPA defocus estimation  error on  defocus distance
(\cite{Blanc-a-03a}).  In   other  words  the  error  on   the  known  defocus
application  directly translates  into an  error on  defocus  estimation. When
introducing  a defocus  diversity,  the  NCPA $a_4$  coefficient  is the  only
polynomial concerned by this bias.

Considering an error  of 10nm on the known defocus, the  error on the measured
$a_4$ is very close to 10nm whereas the total error on the other modes (mainly
spherical aberration and astigmatism) is smaller than 1 nm. The same behavior
was found  when using an  astigmatism as known  aberration. For 10nm  error on
known astigmatism, 10nm error is found on the astigmatism and only 1nm for the
other polynomials in total.

Note that the PCL is not able to compensate for this systematic
bias in the  PD algorithm. The only  way to determine the defocus  is by using
other approaches: e.g.  trying different defocus values to  optimize the image
quality  or using an  other diversity  mode only  for defocus  measurement. In
Section  \ref{exp-res},   we  present   experimental  results  of   these  two
approaches.

\section{Laboratory results}
\label{exp-res}

Both the PCL  iterative compensation method and the various
algorithm optimizations have been experimentally tested on the ONERA AO bench.
It operates with a fibered laser  diode source of 4$\mu$m core size working at
633 nm and located  at the entrance focal plane of the  bench. The laser diode
can be considered as  a incoherent source, since it is used  at very low power
and is therefore weakly coherent with  a large number of modes. The wave-front
corrector includes a  tip-tilt mirror and a 9 by  9 actuator deformable mirror
(69  valid actuators).  The Shack-Hartmann  WFS,  working in  the visible,  is
composed of an 8 by 8 lenslet array (52 in the pupil) and an 128 by 128 pixels
DALSA camera. The WFS sampling frequency  is set to 270 Hz. The imaging camera
is a  512 by 512 Princeton  camera with 4e-/pixel/frame  Read-Out Noise (RON).
The control law used for the AO closed loop is a classical integrator.

An  accurate  estimation  of  image  quality  is  mandatory  to  quantify  the
efficiency   of  the  PCL  and   to  compare   the  different
modifications/improvements of  the PD  algorithm. The Strehl  Ratio (SR)  is a
good  way to  estimate image  quality,  but it  is definitely  not obvious  to
compute  it on  real image  with  a high  accuracy: this  particular point  is
addressed in  appendix with special  care to the  definition of error  bars on
SR estimation.



A  SNR  of $10^4$  in  the focused  image  (see  Equation (\ref{eqn-snr}))  is
sufficient to observe the  5 first Airy rings getting out of  the RON. In this
case,  the PD  estimation  will therefore  be  highly accurate  for the  first
Zernike modes.
  
In order to take advantage of  the regularization and to minimize the aliasing
effect in the measurement, the phase estimation  by PD is done on the 75 first
Zernike polynomials starting at the defocus (from $Z_4$ to $Z_{78}$) and gives
75  Zernike  coefficients (from  $a_4$  to  $a_{78}$).  Nevertheless, we  only
compensate for the first Zernike polynomials (from $Z_4$ to $Z_{28}$), because
of the  limited number of actuators of  the DM. These numbers  of Zernike will
always be used in the next section, except where otherwise stated.

\subsection{Test of the ``pseudo-closed loop'' process}
\label{sec-test_PCL}

For the test of this iterative  method so called pseudo-closed loop (PCL), the
images  used  to  perform PD  are  recorded  with  a  very  high SNR  ($SNR  =
3\times10^4$ for the focused image) so to  be in a noise free regime. This SNR
level corresponds to an  error on the first 25 Zernike of  less than 0.5nm due
to the  noise. The measured  SR before  any compensation is  70\% at
633nm. A  conventional PD algorithm  (without regularization) is  considered here
(because of the high SNR).

\begin{figure}[htbp]
\begin{center}\leavevmode
  \includegraphics[width=\linewidth]{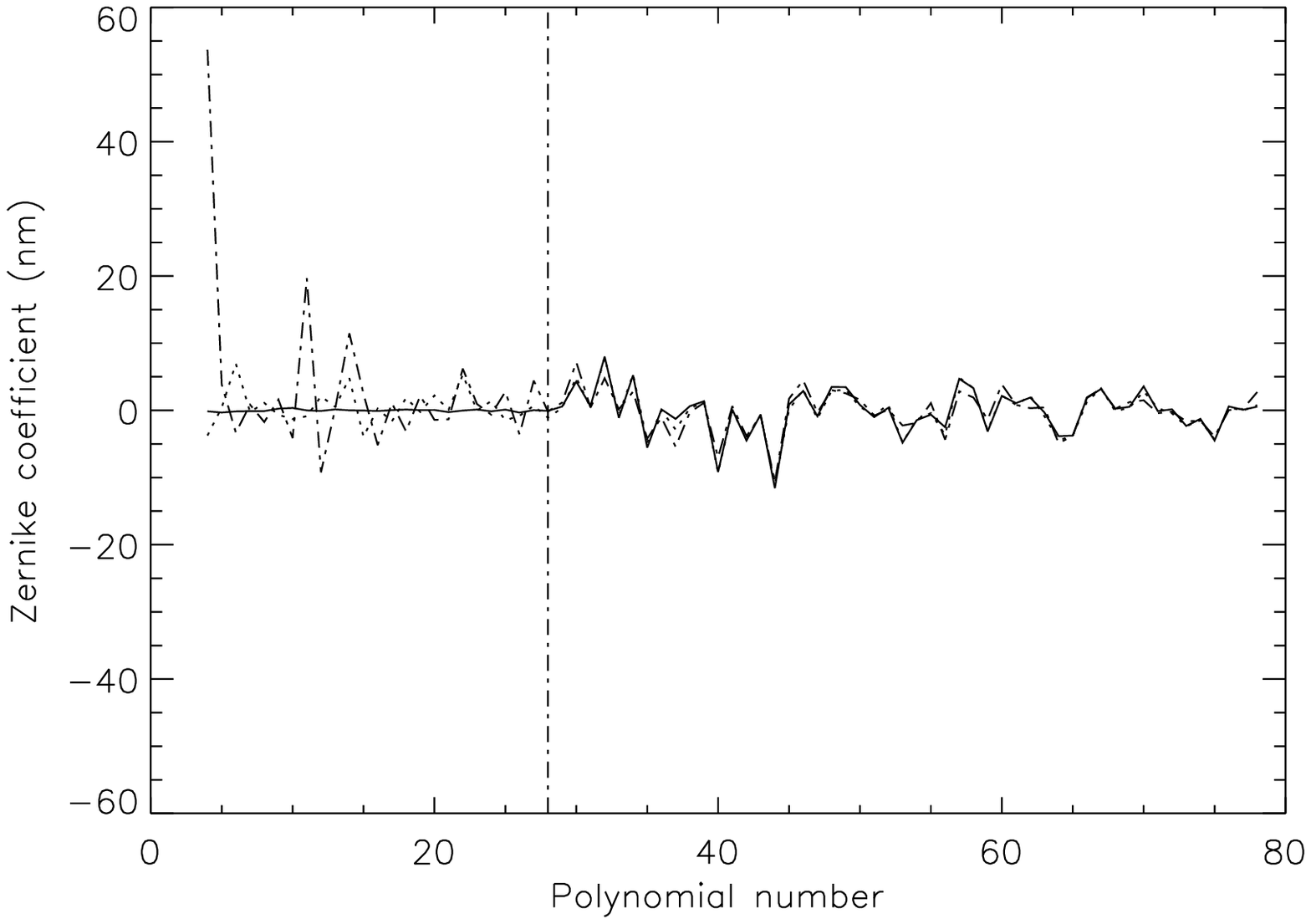}
  \caption{Behavior of  the measured Zernike  coefficients with the  number of
    iteration of  the PCL. Zernike  coefficients up to $Z_{78}$  are estimated
    with conventional PD and SNR=$3 \times 10^4$. Coefficients up to $Z_{28}$ are
    compensated  by PCL  process using  AO closed-loop  with integrator
    control  law. Image wavelength is 632.8nm. Dashed-dotted
    line:  coefficients  before any  compensation.  Dotted line:  coefficients
    after one iteration. Solid line: coefficients after 10 iterations.}
    \label{fig-Z_28_allit}
  \end{center}
\end{figure}

Figure  \ref{fig-Z_28_allit} shows  the  75 Zernike  coefficients measured  at
different iterations of the  PCL procedure. The first iteration
corresponds to the measurement of the NCPA without any pre-compensation.

Only the  first 25 coefficients are  corrected, while 75 are  measured at each
iteration.  The figure shows  the extremely  good correction  of the  25 first
ones, while the 50 higher order modes remain quasi identical.

\begin{figure}[htbp]
 \begin{center}
     \includegraphics[width=\linewidth]{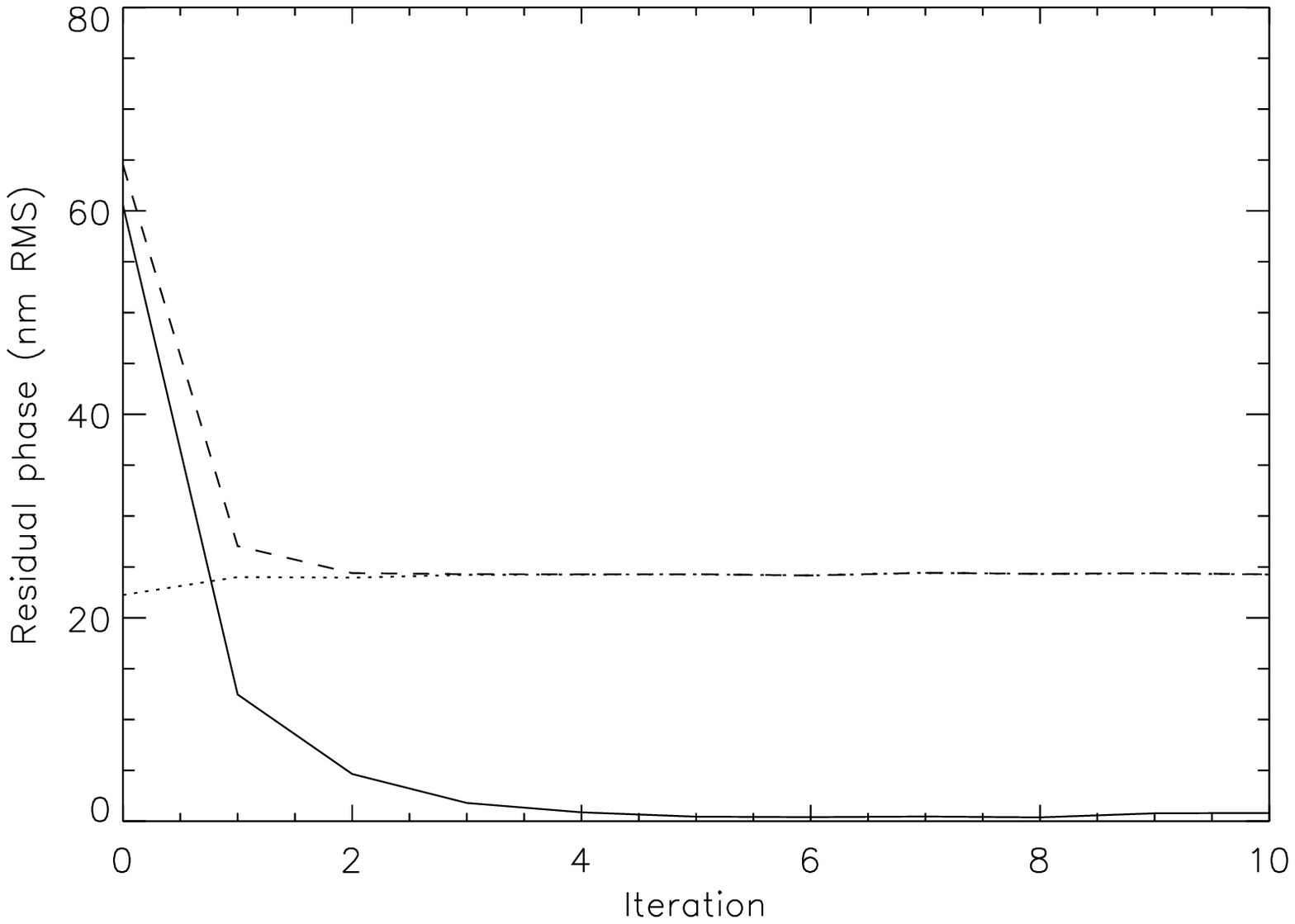}
     \caption{Evolution  of the  residual  error as  a  function of  iteration
       number, for  the corrected [solid  line] and uncorrected  [dotted line]
       Zernike  modes  and for  all  the  measured  modes [dashed  line].  
       Conditions are the same as in Figure \ref{fig-Z_28_allit}.}
      \label{fig-ai_glob}
 \end{center}
\end{figure}

Figure   \ref{fig-ai_glob}  shows   the  evolution   of  the   residual  error
$\sigma=\sqrt{\left(\sum_{k=M}^{k=N}a_k^2\right)}$   for  the  pre-compensated
polynomials  ($M=4$ and $N=28$),  for the  higher order  non-corrected Zernike
modes  ($M=29$ and $N=78$), and for  all the  measured polynomials  ($M=4$ and
$N=78$).  After  4 iterations,  the  global  residual  phase computed  on  the
corrected Zernike modes ($M=4$ and $N=28$) is lower than 1 nm RMS (not limited
by  noise  in  the  images)   whereas  the  residual  phase  computed  on  the
non-corrected  Zernike  ($M=29$  and  $N=78$) modes  remains  quasi  identical
passing from  22 to around  24 nm RMS.  After convergence, the  total residual
error on the 78 first Zernike polynomials is 24 nm.

Finally, for  each iteration, a  SR value ($SR_{Im}$)  can be measured  on the
focused  image. In  addition another  SR value  ($SR_{Zern}$) can  be computed
using  the   coefficients  estimated  by   the  PD  algorithm   (see  Equation
\ref{eqn-phase_trunc} in Appendix). We compare in Figure \ref{fig-strehl_comp}
the  measured  $SR_{im}$  and  the  estimated $SR_{Zern}$  as  a  function  of
iteration number. Both  $SR_{im}$ and $SR_{Zern}$ have the  same behavior. The
maximum value  achieved by $SR_{im}$ is  93.8\% at 633nm.  After 2 iterations,
$SR_{Im}$ reaches a convergence plateau.

\begin{figure}[htbp]
  \begin{center}
    \includegraphics[width=\linewidth]{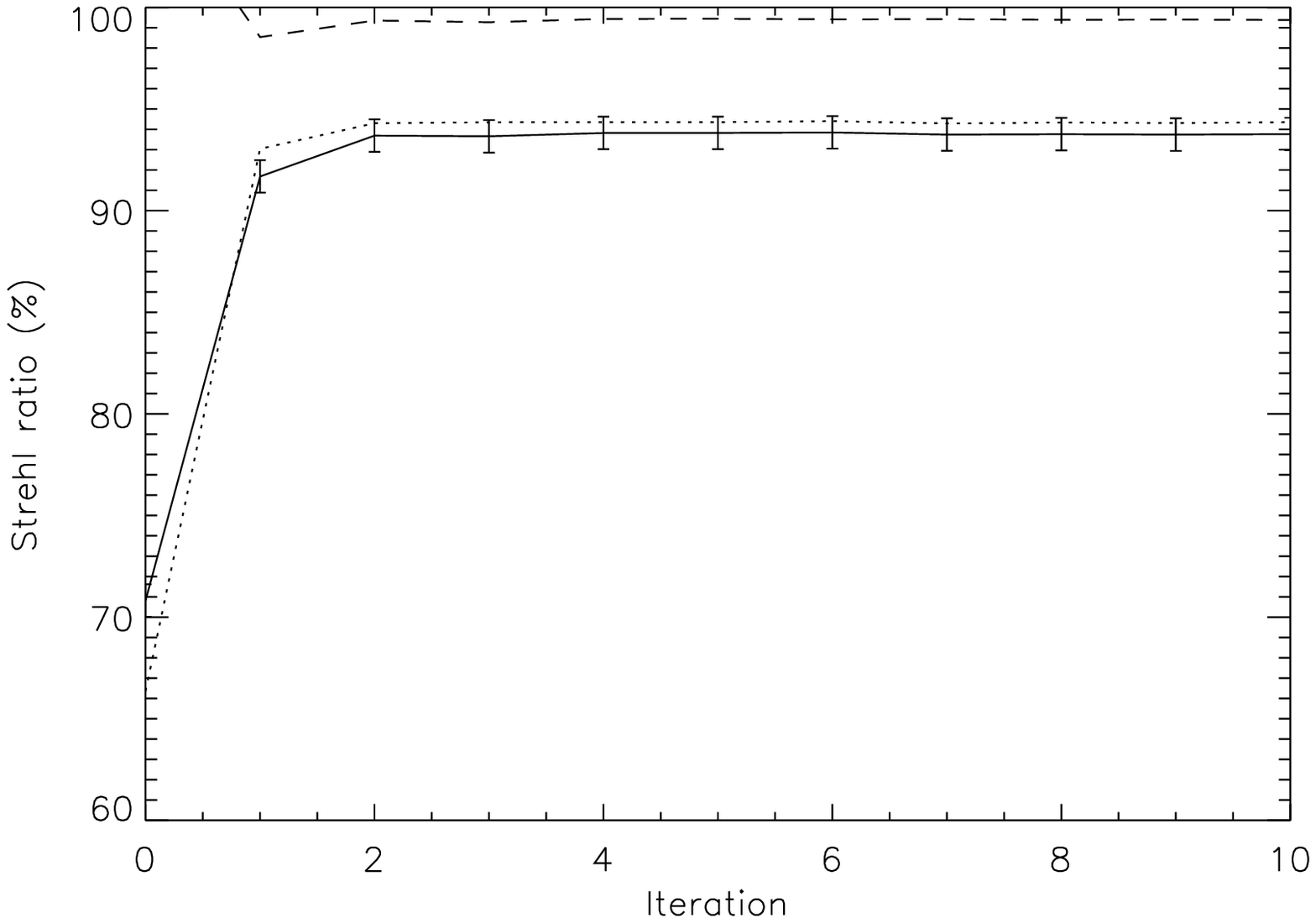}
    \caption{Evolution of  SR with iteration number.  SR is measured
      on the focal plane images ($SR_{im}$). The
      SR computed  from the measured  NCPA ($SR_{Zern}$) is plotted  in dotted
      line.  SR bias is  uniform  and estimated to  0.008.  The ratio
      between $SR_{im}$ and $SR_{Zern}$  is plotted in dashed line. Conditions
      are the same as in Figure \ref{fig-Z_28_allit}.}
      \label{fig-strehl_comp}
  \end{center}
\end{figure}

\begin{figure}[htbp]
  \begin{center}
    \includegraphics[width=\linewidth]{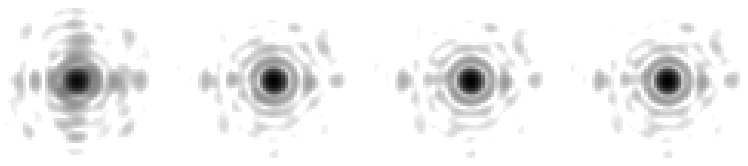}
    \caption{Focused images  obtained on ONERA AO  bench (logarithmic
      scale)    corresponding   to    the   4    first   points    of   Figure
      \ref{fig-strehl_comp}. The one on the left is the image obtained without
      any pre-compensation.  The last image  SR is 93.8\%. Conditions  are the
      same as in Figure \ref{fig-Z_28_allit}.}
      \label{fig-psfx}
  \end{center}
\end{figure}

We plot  on the same figure  the ratio between $SR_{im}$  and $SR_{Zern}$. The
difference  between  $SR_{im}$  and   $SR_{Zern}$  can  be  explained  by  the
unestimated  high  order  coefficients  (higher  than  $a_{78}$)  and  the  SR
measurement  bias due  to  uncertainty on  the  system (exact  over-sampling
factor,  background subtraction precision,  exact fiber  size and  shape, see
Appendix for more details).  This ratio $\frac{SR_{im}}{SR_{Zern}}$ is roughly
constant  after  the first  iteration  and its  value  at  convergence can  be
estimated to 99.4\% which corresponds to a 8 nm RMS phase error. The different
values for the first iteration are explained by the approximation of SR by the
coherent energy in $SR_{Zern}$, which is only valid for small phase variance.

We plot  in Figure  \ref{fig-psfx} the focused  images recorded on  the camera
without NCPA pre-compensation and after  respectively 1, 2 and 3 iterations of
the PCL scheme. The  correction of low order aberrations allows
for the cleaning of  the center of the image where two  Airy rings are clearly
visible  with the first  one being  complete. Around  these rings,  we observe
residual speckles due to the uncompensated higher order aberrations.

\subsection{Defocus determination}
\label{sec-defoc_optim}

As explained  in Section  \ref{sec-PD-DM}, NCPA defocus  estimation has  to be
considered with  a particular care  since it is  biased by uncertainty  on the
``known  aberration'' (actually  not  perfectly known)  introduced  in the  PD
method.  In order to  overcome this  bias, two  approaches are proposed
after the convergence of the pre-compensated scheme. The first one is based on
a  SR  optimization,  the  second  one  is  a  one-shot  measurement  with  an
``astigmatism'' phase-diversity.

\subsubsection{SR optimization}
\begin{figure}
\begin{center}\leavevmode
  \includegraphics[width=\linewidth]{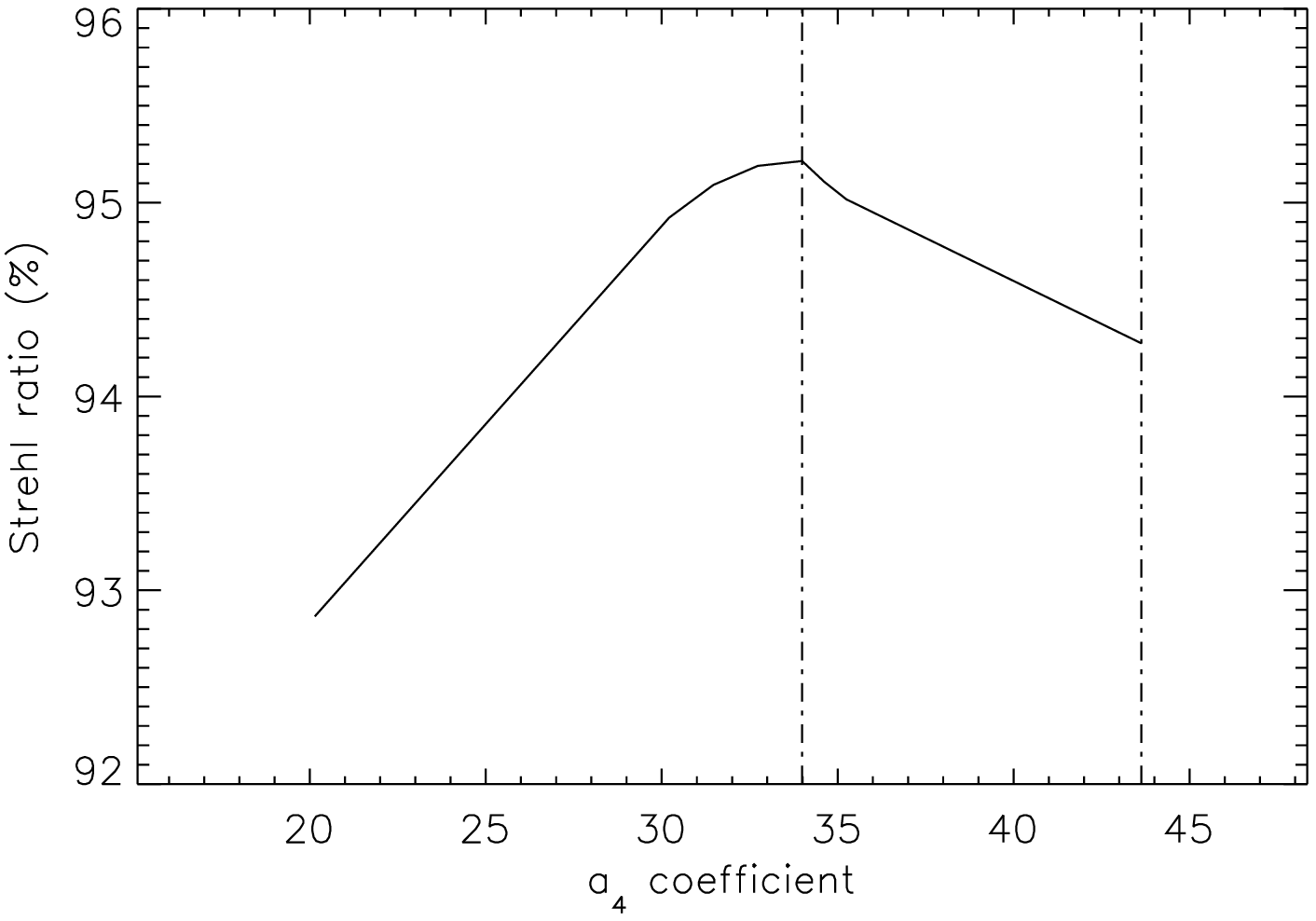}
  \caption{Optimization of the coefficient $a_4$, by  changing its
    value before applying the correction slopes. $a_4=43nm$ is the value estimated
    by PD and $a_4=34nm$ is the  value estimated by the maximum SR.
    Conditions are the same as in Figure \ref{fig-Z_28_allit} except for the
    number of compensated Zernike : up to $Z_{36}$.}
  \label{fig-defoc_optim}
  \end{center}
\end{figure}

After  a  few  iterations of  the  PCL  (enough to  reach  the
convergence), we modify the pre-compensated $a_4$ coefficient in a given range
around the estimated  coefficient and we measure the  corresponding SR. Figure
\ref{fig-defoc_optim} shows the SR evolution with the value of $a_4$.
 The maximum SR is obtained for $a_4=34nm$ which is somewhat different from
the value given by PD (43nm). The resulting gain in term of SR is around 1\%.

\subsubsection{Astigmatism phase-diversity}
An alternative way to perform the  NCPA defocus optimization is to use another
known aberration  between the two images.  As explained in  the first section,
defocalisation is  generally used because of its easy implementation.
In  our case  the  deformable mirror  itself  is used  to  generate the  known
aberration. Thus any Zernike polynomial can be considered, as long as it is an
even radial order (to solve the estimation phase indetermination) and feasible
by the DM.

At  convergence  of  the  PCL,  we  acquire a  pair  of  images  differing  by
astigmatism ($Z_5$). The PD  measurement performed gives coefficients $a_4$ to
$a_{78}$, with a  biased estimation of $a_5$ (the previous  error due to model
uncertainty is now done on $a_5$ instead of $a_4$) while the coefficient $a_4$
is  now correctly  estimated. The  value  given for  $a_4$ by  this method  is
$a_4=35nm$ which is fully compatible  with the SR optimization of the previous
section.

\subsection{Test of optimized algorithms}

Let  us now  experimentally validate  the various  PD  algorithm modifications
proposed in Section  \ref{sec-opt} (that is non-uniform  noise model and phase
regularization).

In order to  experimentally test these improvements, two  different regimes of
SNR have been used: high SNR  regime as before, $SNR=3\times 10^4$ and low SNR
regime, $SNR=10^2$ obtained with the smallest exposure  time while acquiring
the pair of images.

Table \ref{tab-results} gathers the various  SR values obtained after
convergence  of the  PCL  process for  the  different algorithm
modification.

\begin{figure}[htbp]
\begin{center}\leavevmode
  \includegraphics[width=\linewidth]{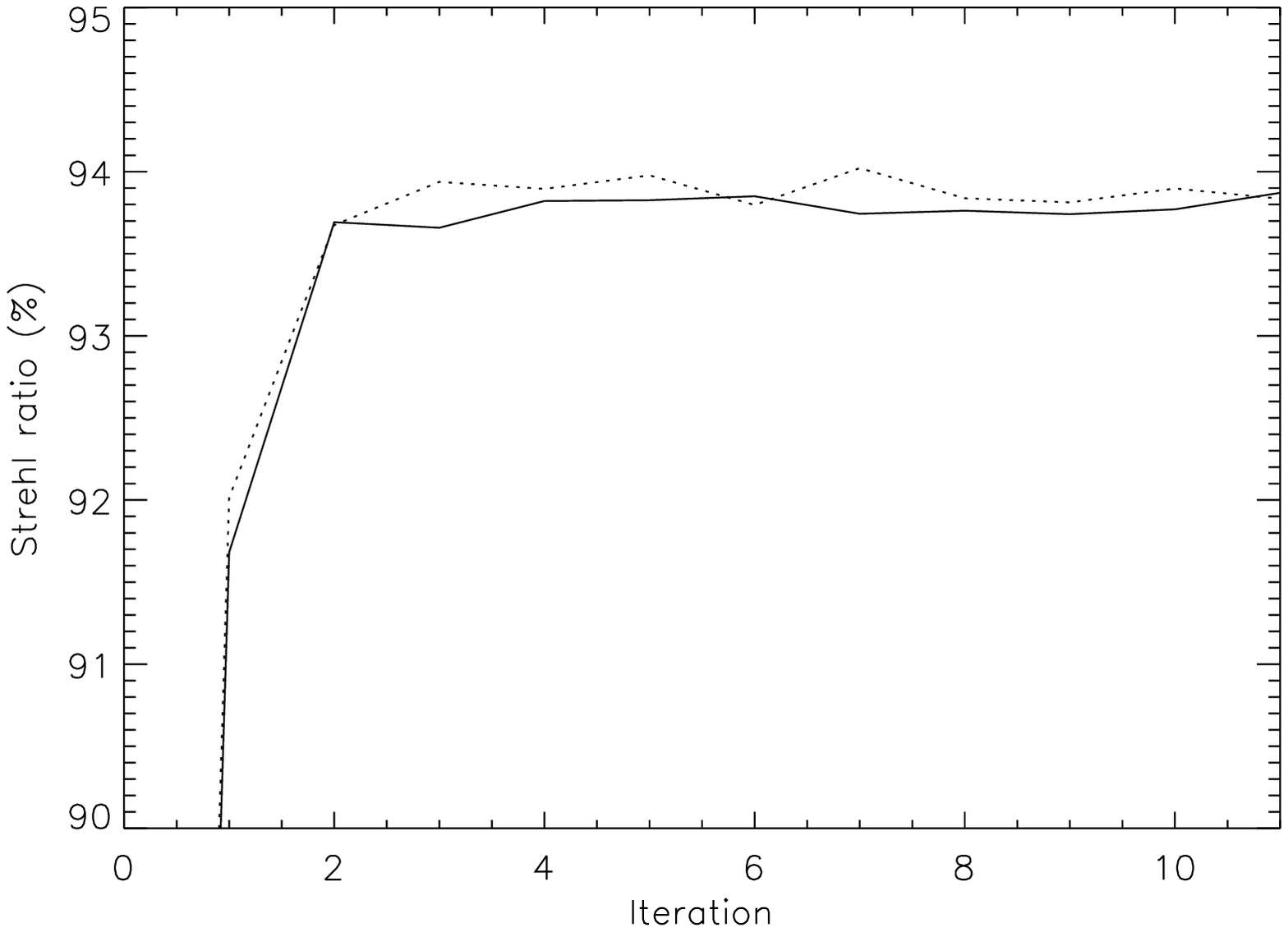}
  \caption{Evolution of SR with iteration number. Solid  line : the
    conventional  algorithm (uniform noise  model). Dotted  line :  the
    non-uniform  noise  model algorithm. The  measurements  were done at the
    time for each iteration.  Conditions are  the  same as  in Figure  \ref{fig-Z_28_allit},
    except for the used PD algorithm.}
  \label{fig-Z_28_u_nu}
  \end{center}
\end{figure}

At  high SNR,  the correction  remains extremely  good whatever  the algorithm
configuration. The gain brought by the non-uniform noise model is rather small
and  the limitation  comes from  other error  terms,  especially non-corrected
modes.  However, there  is  a  slight gain  of  0.1\% of  SR  shown in  Figure
\ref{fig-Z_28_u_nu} which was predicted by the theory. At low SNR, the gain is
much higher, 10\% in SR but this is  the result of a one-shot test, not a mean
gain obtained on a large number of trials.

The gain brought by the use of  the phase regularization term is shown in low SNR
regime. At  $SNR=10^2$, the conventional algorithm  without regularization barely
estimates the phase. It leads to a poor  result after pre-compensation:~the
saturation plateau  remains around 72.1\%  showing no real improvement  on the
image. When the regularization term is  added, the SR value reaches 91.9\%. 
The NCPA  are estimated  at almost the  same accuracy than in  high SNR
regime.

As  shown in  section \ref{subsec-sim},  the use  of regularization  allows the
minimization of  the noise amplification  on the  high order  estimated modes  and 
enhances the estimation  accuracy of the lower orders.  This properties will be
particularly useful with an infrared camera where the SNR in the image could
be limited and when a large number of modes have to be estimated. Note also
the substantial gain brought by the use of the non-uniform noise model
algorithm when compared to the conventional one and also when coupled to the
phase regularization.

\begin{table}
\caption{SR obtained with the different optimized algorithms.}
\begin{center}
  \begin{tabular}{|c|c|c|c|c|} \hline
    &Conventional &Non-uniform & Phase &Regularization and \\
    &algorithm&noise model & Regularization &non-uniform noise model \\\hline 
    
    Max SR&  \multirow{2}{*}{93.8\%}  &\multirow{2}{*}{93.9\%}  & \multirow{2}{*}{93.8\%} & \multirow{2}{*}{93.9\%} \\
    for high SNR  & & & & \\ \hline
    
    Max SR &\multirow{2}{*}{72.1\%}  &\multirow{2}{*}{81.2\%}  & \multirow{2}{*}{91.9\%} & \multirow{2}{*}{92.3\%} \\  
    for low SNR   & & & & \\ \hline  
  \end{tabular}
\end{center}
    \label{tab-results}
\end{table}

\subsection{Number of modes}


An additional  test has  been performed with  the PCL.  In this
test, the conditions are the same than previously, i.e. : 75 Zernike modes are
measured by PD. The SNR in the images used by PD is very high (10$^4$). The PD
algorithm used to  perform NCPA measurement is the  conventional one. The NCPA
$a_4$ estimation is unbiased (see Section \ref{sec-defoc_optim}). In this test
42 Zernike modes  have been compensated, instead of  25. 

\begin{figure}[htbp]
  \begin{center}\leavevmode
     \includegraphics[width=0.3\linewidth]{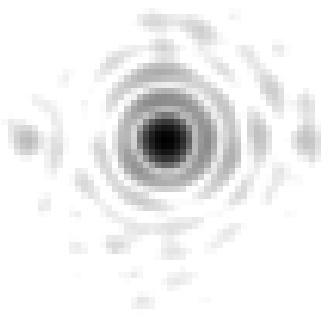}
     \caption{PSF obtained after three iterations of NCPA pre-compensation. 42
       modes are compensated (from $Z_4$  to $Z_{45}$), and the exact value of
       $a_4$ has been optimized. SR is 98.7\%, $\lambda=632.8nm$.}
     \label{fig-max}
  \end{center}
\end{figure}

\begin{figure}[htbp]
   \begin{center}\leavevmode
     \includegraphics[width=\linewidth]{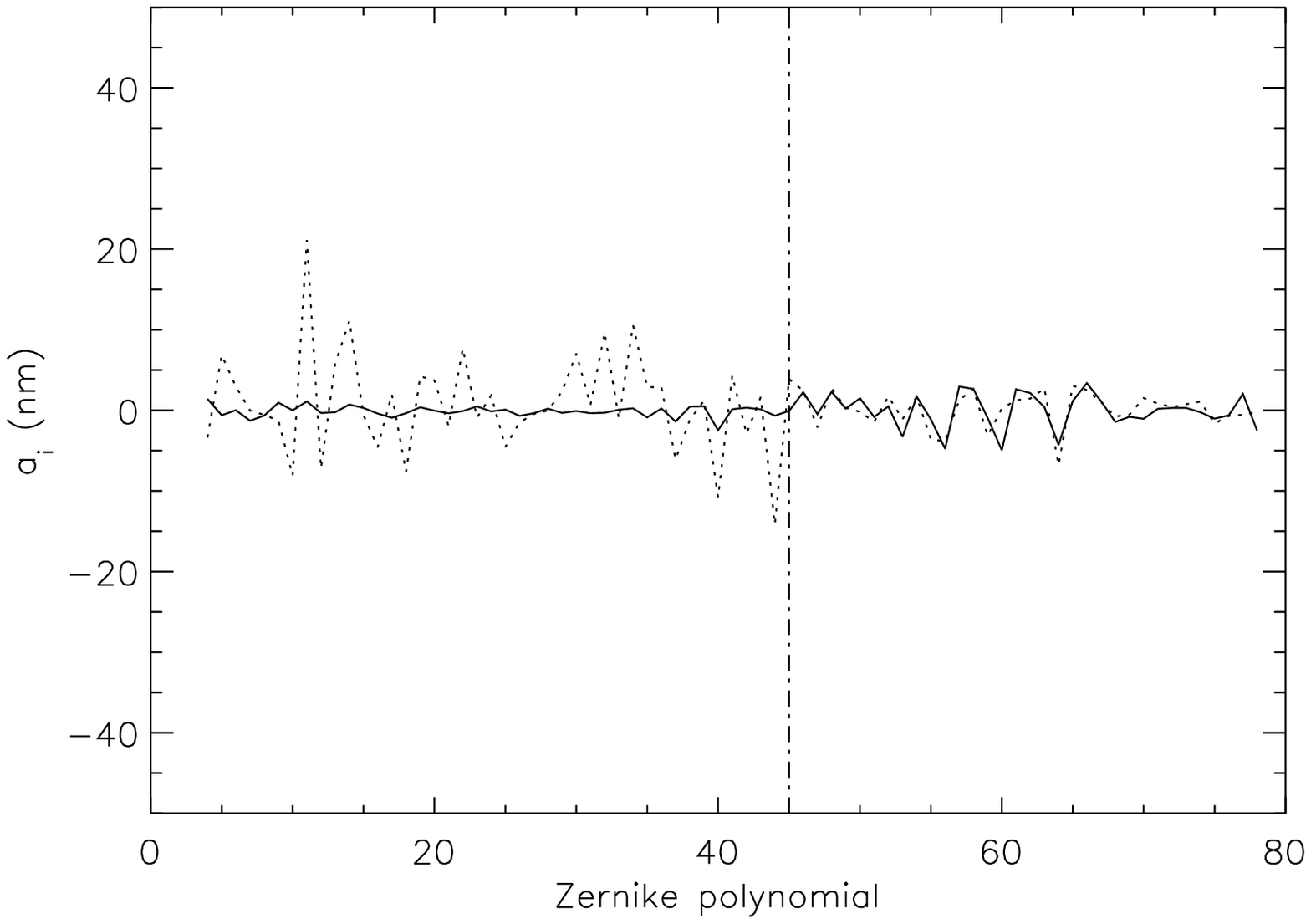}
   \caption{Zernike coefficients measured before compensation (dotted line) and after 3
     iterations of PCL (Dashed line, corresponding to the PSF shown in the Figure
     \ref{fig-max}). The conditions are the same than in Figure \ref{fig-max}.}
   \label{fig-ai_max}
  \end{center}
\end{figure}

The corresponding PSF is shown in Figure \ref{fig-max}, revealing up to 4 Airy
rings.  A SR  of  98.7\%  is obtained.  Figure  \ref{fig-ai_max} presents  the
measured  Zernike  coefficients  before   any  compensation  and  after  three
iterations  of  PCL corresponding  to  Figure  \ref{fig-max}.  A 13  nm  total
residual error  was estimated  on the 75  Zernike coefficients,  a substantial
gain when compared to the result  of Figure \ref{fig-ai_glob} (24.5 nm). Up to
$Z_{45}$, the  residual error  is 4.5nm, while  the higher  order contribution
remains below than  12.5 nm. We observe that on the  first order, the residual
error  is   slightly  higher   when  compared  to   the  results   of  Section
\ref{sec-test_PCL}     (only     1nm).     As     explained     in     section
\ref{sec-number_compensated_modes}, the  high order Zernike modes  of NCPA are
not well  fitted by  the DM.  That induces some  coupling effects  between the
compensated  highest   order  modes  (from  $Z_{29}$  to   $Z_{45}$)  and  the
uncompensated  ones (above  $Z_{45}$). They  induced some  aliasing  effect on
lower   modes,   slightly   decreasing  their   pre-compensation   efficiency.
Nevertheless,  the gain  brought by  the partial  correction of  highest order
modes is higher than the loss due to aliasing effects.

Note that these performance were obtained using a Kalman filter in the AO loop
as developed  for optimized compensation  of the turbulence\cite{Petit-p-06},
and not a simple integrator corrector  as for the previous result. This Kalman
filter  uses  the  first 130  Zernike  modes  for  the WFS  phase  regularized
estimation, leading  to partially overcome  the limitations linked to  the bad
fitting of  the high order Zernike  by the DM.  It was not possible  to obtain
such a high SR (98.7\%) with the integrator in the same conditions.

\section{Discussion}

We discuss here the gain brought  by our new approach for NCPA measurement and
compensation on extreme AO systems  for extrasolar planet detection. This type
of instrument  requires very high AO  performance in order  to directly detect
photons  coming from  very faint  companions  orbiting their  parent star.  An
example of  such an extreme AO system  applied to direct exoplanet  detection is SAXO
\cite{Fusco-a-06b}, the  extreme AO system of SPHERE \cite{Beuzit-p-05}.  SPHERE is a
2nd  generation VLT instrument  considered for  first light  in 2010.  It will
allow one to detect hot Jupiter-like planets with the contrast up to $10^{6}$.

Characteristics  of SAXO are  the following  \cite{Fusco-a-06b} :  a 41  by 41
actuator DM and a WFE  budget of 80nm after extreme AO correction (90\%  SR in H band).
The specification of SAXO is to  compensate for NCPA with at least 100 Zernike
modes (goal 200) and to allocate a 8nm WFE on these modes.

The  high number  of  actuators available  on the  DM  allows one  to fit  the
requested  number of  modes (no  fitting problem).  Therefore, the  only error
source is therefore assumed to be a PD estimation error.

According to Figure \ref{fig-regul_simu} and  to the fact that precision of PD
is  inversely proportional  to SNR  in image  \cite{Meynadier-a-99}, a  SNR of
$10^4$  is sufficient to  have a  precision of  $0.4nm^2$ on  each of  the 100
measured Zernike,  i.e. a  residual WFE of  6nm. This  SNR can be  achieved by
averaging a number  of images recorded by the  infrared camera. Moreover, this
level of  residual correction  per Zernike mode  is fully compatible  with the
results  obtained  on  our  AO  Bench  (see  Figure  \ref{fig-Z_28_allit}  and
\ref{fig-ai_max}). The  goal of  100 Zernike for  compensation and 8nm  of WFE
after compensation  is therefore fully  achievable with the PCL  and optimized
algorithms presented here.

\section{Conclusion}

We  have  proposed  and  validated  a  new  and  efficient  approach  for  the
measurement and pre-compensation  of the NCPA. First, the  measurement quality
of NCPA
has been improved via the  optimization of PD algorithm (accurate noise model,
phase regularization). Moreover the  limitation imposed by model uncertainties
during the  pre-compensation process  has been  overcome by the  use of  a new
iterative approach (PCL).

We have  experimentally validated this  new tool on  the ONERA AO  bench. Very
high SR has  been obtained (around 98.7\% of  SR at 633 nm, that  is less than
14nm of residual defects). The   residual WFE on corrected modes is less than
4.5nm RMS. We have estimated  the residual error on uncorrected aberrations to
be  less  than  12.5nm  RMS.   Solutions  have  been  proposed  to  deal  with
experimental   issues   (such   as   defocus  uncertainty)   of   the   PD
implementation.

The experimental results  presented in this paper allow us  to be confident in
our  capability   of  achieving  the  challenging   performance  required  for
extrasolar planet  direct detection.  The residual errors  obtained on  our AO
bench for NCPA  compensation are fully compatible with the  error budget of an
extreme AO system like SAXO. Using 100  Zernike in the compensation we should achieve
a 8nm residual error, corresponding to 99.9\% SR, @1.6$\mu m$.

\appendix
\label{app-strehl}

\section{Strehl ratio estimation}

\subsection{Strehl Ratio estimation in focal plan images}

A  widely  used  performance  estimator   in  AO  is  the  SR.  Its
experimental estimation is  difficult  and requires optimized algorithms
able to deal  with a number of experimental biases or  noises. We propose here
an  efficient SR measurement  procedure. The  SR  is defined  as the
ratio of the  on-axis value (or tilt free) of the  aberrated image $i_{ab}$ on
the on-axis value of the aberrations-free image $i_{Airy}$. Several parameters
have to be taken into account in order to obtain accurate and unbiased values:
the residual  background and the noise in  the image, the CCD  pixel scale, and
the  size of the calibration source used to calibrate the NCPA. 

\begin{figure}[htbp]
   \begin{center}\leavevmode
   \includegraphics[width=\linewidth]{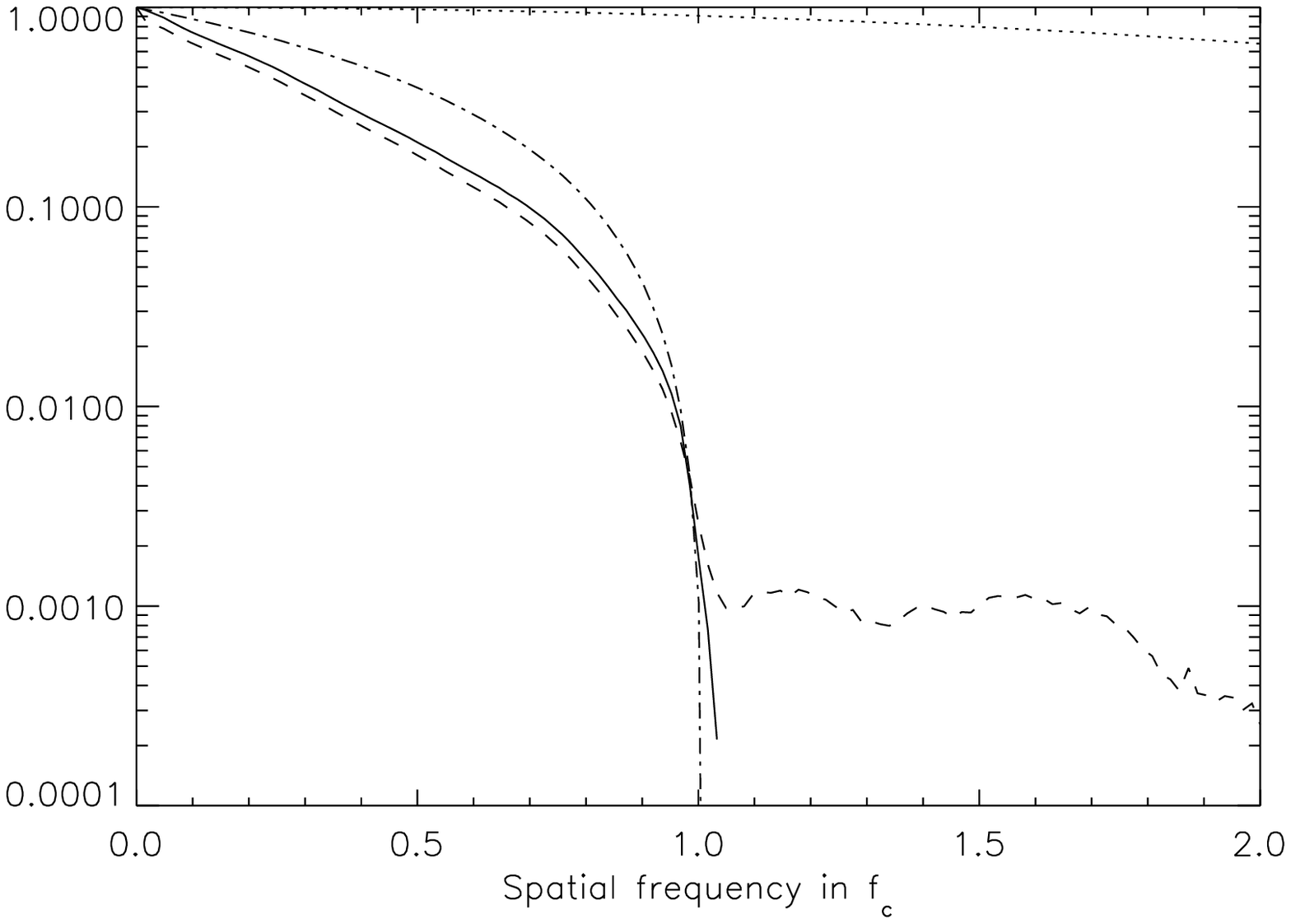}
   \caption{Measured OTF  from the image (Dashed line),  the OTF corrected for noise  and background
     contributions (solid line), and the adjusted Airy OTF as obtained by the SR measurement
     procedure (dot-dashed line. The cut-off frequencies are adjusted to be superimposed. The
     transfer function of the CCD is also given (dotted line).}
   \label{fig-strehl_measure}
  \end{center}
\end{figure}

The SR value can be computed using the following equations: 

\begin{equation}
  SR_{im}= \frac{i_{ab}(\vec{0})}{i_{Airy}(\vec{0})}
        = \frac{\int \tilde i_{ab} (\vec{f}).d\vec{f}}{\int \tilde i_{Airy}(\vec{f}).d\vec{f}}
  \label{eqn-strehl}
\end{equation}
where $\tilde i_{ab} $ and $\tilde i_{Airy}$ are the optical transfer function
(OTF) of  the aberrant system  and of the aberration-free  system respectively
($\tilde  i$ standing  for Fourier  Transform  of $i$),  $\vec{f}$ a  position
variable in Fourier space. We developed  a procedure calculating the SR in the
Fourier domain. Considering OTF rather than the PSF for SR estimation presents
several   advantages  which   are   summarized  below,   and  illustrated   in
Figure~\ref{fig-strehl_measure}.

\begin{itemize}
\item First, an analysis of the FT  of the aberrated image (OTF) allows a fine
  subtraction  of the  residual  background  which can  be  estimated from  a
  parabolic fit at  the lowest frequencies of the  aberrated OTF excluding the
  zero frequency.
\item  An  important   point  is  the  adjustment  of   the  cut-off  frequency
  ($f_c=\frac{D}{\lambda}$)  in  $\tilde i_{Airy} $  (value in  frequency  pixel
  directly linked to  the image pixel scale) to the  experimental value in the
  aberrated OTF. The  pixel scale is also used by PD to estimate the
  phase.
\item Actually,  all the OTF values  for frequencies greater  than the cut-off
  frequency  are  only  noise.  An   estimation  of  this  level  and  then  a
  subtraction to the  aberrated OTF allow us to refine  the SR estimation. In
  Figure~\ref{fig-strehl_measure},  for  higher  frequencies  than  $f_c$  the
  aberrated OTF presents a noise plateau at 1.e-3.
\item At last the procedure also  takes into account the transfer function of
  the CCD and of the FT of the object, when the last one is partially resolved
  by the optical system.
\end{itemize}

\subsection{Errors in Strehl ratio estimation}

Practical  instrumental limitations  degrade the  SR estimation  accuracy even
when using an optimized algorithm as in Section \ref{sec-opt}. It is important
to quantify their influence in order to give error bars on SR values.

\subsubsection{Influence of residual background}
\label{sec-SR_b}

Let's first  study the influence of  a residual uniform  background $\delta B$
per pixel on the estimation of the $SR$ in an image $i_f$.

Using Equation \ref{eqn-strehl} we  can express the background contribution
in the SR  computation as follows: 

\begin{eqnarray}
  \label{eqn-sr_im}
  SR_{im}&=&\frac{\left(i_f(\vec 0)+\delta B\right)\int{i_{Airy}(\vec\alpha)d\vec\alpha}}
  {i_{Airy}(\vec 0)\int{\left(i_{f}(\vec\alpha)+\delta B\right)d\vec\alpha}} \\ \nonumber
  &\simeq& SR\left(1+\frac{\delta B}{i_f(\vec 0)}\right)
  \left(1-\frac{N^2\delta B}{\int{i_{f}(\vec\alpha)d\vec\alpha}}\right) \\ \nonumber
\end{eqnarray}

Note   that  Equation   \ref{eqn-sr_im}  takes   into  account   the  required
normalization      by      the      total      flux     in      the      image
($\int{\left(i_{f}(\vec\alpha)+\delta B\right)d\vec\alpha}$),  where the image
$i_{f}(\vec\alpha)$          has          $N$x$N$          pixels          and
$\int{i_{Airy}(\vec\alpha)d\vec\alpha}=1$. Considering  that generally $\delta
B <<  i_f(\vec 0)$, a  residual background $N^2\delta  B$ equal to 1\%  of the
total  flux in  the  image modifies  the SR  value  of 1\%  as well.  The
residual  background  in  our  images  after subtraction  of  the  calibrated
background and after  correction by fitting of the  lowest frequencies of the
OTF is  estimated at $\pm 0.1  \%$ of the  total flux using images  of 128x128
pixels. The SR estimation accuracy is therefore  $\pm 0.1\%$.

\subsubsection{Influence of an uncertainty on the pixel scale}
\label{sec-SR_e}

The pixel scale is a parameter to be experimentally estimated since it greatly
depends on components characterization  and system implementation. It plays an
important  role in  the  NCPA estimation  through  the PD  algorithm. It  also
impacts the SR estimation quality. Its influence is therefore major on the
whole procedure discussed in this paper. In the next section, we emphasize its
influence on the SR measurement.

Assuming that the OTF profile for an Airy pattern has a linear shape, a simple
computation shows that the relative SR modification $\frac{\delta SR}{SR}$
is directly equal to twice  the relative precision $\frac{\delta e}{e}$ on the
pixel scale $e$, that is: 

\begin{equation}
  \frac{\delta SR}{SR}=-2\frac{\delta e}{e}
\end{equation}

It  is  clear  that knowledge  of  $e$  is  essential  to obtain  an  accurate
estimation of SR. Nevertheless, $e$ value doest not evolve with time Therefore
the estimation error on SR is constant for all the test. The impact on SR is a
bias. In  other words, if $e$  value is critical for  absolute SR computation,
its influence is  dramatically reduced when only relative  evolution of SR are
considered.  (Gain brought  by a  new  approach of  NCPA pre-compensation  for
example).

Now the  essential question is  : ``with which  accuracy do we know  the pixel
scale?''. In images taken on different days, the measured cut-off frequency is
stable  with an  uncertainty lower  than half  a frequency  pixel.  The random
relative error on the pixel scale  is 0.4\% for 128 frequency pixels. Finally,
we used the same measured pixel scale ($e=\lambda / 4.1 D$) in data processing
of all the performed experiments (NCPA measurement and SR estimation). The
relative  error on SR  estimation is  therefore a  bias of  0.8\%. Because
$SR\simeq100\%$, the absolute error on $SR$ is  $0.8\%$.

\subsubsection{SR accuracy in experimental data}
It is  now possible  to estimate  a global accuracy  of the  SR estimation
using   the   results    of   Section   \ref{sec-SR_b}   and   \ref{sec-SR_e}:
$\sigma_{SR}=\sqrt{0.8^2+0.1^2}\simeq.81\%$. The pixel scale bias is the main
contribution to this value.

\subsection{SR estimated using the measured Zernike coefficients}

Another  way  to  estimate  $SR$   is  to  use  the  residual  phase  variance
$\sigma_{\phi}^2$       to       compute       the       coherent       energy
$e^{\left(-\sigma_{\phi}^2\right)}$.  The  residual   phase  variance  can  be
directly obtained  using the PD  estimated Zernike coefficients.  We therefore
define the approximated SR $SR_{Zern}$ by:
\begin{equation}
  SR_{Zern}= \exp\left(-\sum_{k=2}^{N_{max}}a_k^2\right)
  \label{eqn-phase_trunc}
\end{equation}
where $a_k$  stands for  the k$^{th}$ Zernike  coefficient. In one  hand, this
expression of  $SR_{Zern}$ is  a lower  bound for the  true SR  for relatively
large  residual phase  (low SR).  In  the other  hand, $SR_{Zern}$  is a  good
approximation of SR for small residual  phases (high SR). But it slightly over
estimates  it since  $SR_{Zern}$   only  accounts for  the $N_{max}$  first
Zernike  (Equation \ref{eqn-phase_trunc}). Figure  \ref{fig-strehl_comp} shows
the image measured SR ($SR_{im}$) and computed one ($SR_{Zern}$). We verify on
this figure the behavior described here above.

\providecommand{\inpreparationname}{en pr\'eparation}
  \providecommand{\submittedname}{soumis}
  \providecommand{\acceptedname}{accept\'e pour publication}
  \providecommand{\tobepublishedname}{\`a para\^{\i}tre}
  \providecommand{\contractname}{Contrat}
  \providecommand{\conferencedatename}{Date conf\'erence~: }
  \providecommand{\patent}[2]{Brevet #1 #2}
  \providecommand{\firstabbrevname}{1\textsuperscript{\`ere} }
  \providecommand{\secondabbrevname}{2\textsuperscript{\`eme} }
  \providecommand{\thirdabbrevname}{3\textsuperscript{\`eme} }
  \providecommand{\fourthabbrevname}{4\textsuperscript{\`eme} }
  \providecommand{\fifththabbrevname}{5\textsuperscript{\`eme} }
  \providecommand{\sixththabbrevname}{6\textsuperscript{\`eme} }

\end{document}